\documentclass[pre,aps,floats,floatfix,superscriptaddress,nofootinbib,showpacs]{revtex4}
\usepackage[english]{babel}
\usepackage{amsmath,amssymb,amsfonts}
\usepackage[dvips]{graphicx}
\usepackage{amsmath}
\usepackage{fancyhdr}

\usepackage{bm}
\newcommand{\be}{\begin{equation}}
\newcommand{\ee}{\end{equation}}

\begin{document}

\newcommand{\sigmagorro}{\hat{\boldsymbol{\sigma}}}
\title{Dynamics of Annihilation I : Linearized Boltzmann Equation and 
Hydrodynamics}

\author{Mar\'ia Isabel Garc\'ia de Soria}
\affiliation{Universit\'e Paris-Sud, LPTMS, UMR 8626, Orsay Cedex, F-91405 and
CNRS, Orsay, F-91405}
\author{Pablo Maynar}
\affiliation{Laboratoire de Physique Th\'eorique (CNRS
  UMR 8627), B\^atiment 210, Universit\'e Paris-Sud, 91405 Orsay cedex,
  France}
\affiliation{F\'{\i}sica Te\'{o}rica, Universidad de Sevilla,
Apartado de Correos 1065, E-41080, Sevilla, Spain}
\author{Gr\'egory Schehr}
\affiliation{Laboratoire de Physique Th\'eorique (CNRS
  UMR 8627), B\^atiment 210, Universit\'e Paris-Sud, 91405 Orsay cedex, France}
\author{Alain Barrat}
\affiliation{Laboratoire de Physique Th\'eorique (CNRS
  UMR 8627), B\^atiment 210, Universit\'e Paris-Sud, 91405 Orsay cedex, France}
\author{Emmanuel Trizac}
\affiliation{Universit\'e Paris-Sud, LPTMS, UMR 8626, Orsay Cedex, F-91405 and
CNRS, Orsay, F-91405}

\date{\today }

\begin{abstract}
We study the non-equilibrium statistical mechanics of a system of 
freely moving particles, in which binary encounters lead either to
an elastic collision or to the disappearance of the pair. Such a
system of {\em ballistic annihilation}
therefore constantly looses particles. The dynamics of perturbations
around the free decay regime is investigated from the spectral
properties of the linearized Boltzmann operator, that characterize
linear excitations on all time scales. The linearized Boltzmann
equation is solved in the hydrodynamic limit by a projection
technique, which yields the evolution equations for the relevant
coarse-grained fields and expressions for the transport coefficients.
We finally present the results of Molecular Dynamics simulations that
validate the theoretical predictions.

\end{abstract}


\maketitle

\section{Introduction}

Understanding the differences and similarities between a flow of
macroscopic grains and that of an ordinary liquid is an active field
of research \cite{g05,d07b}. From a fundamental perspective, it is
tempting to draw a correspondence between the grains of the former and
the atoms of the latter in order to make use of the powerful tools of
statistical mechanics to derive a large scale description for the
various fields of interest, such as the local density of grains.  A
key difference between a granular system and an ordinary liquid is
that collisions between macroscopic grains dissipate energy, due to
the redistribution of translational kinetic energy into internal
modes.  This simple fact has far reaching consequences
\cite{bte05,d07b}, but also poses an {\it a priori} serious problem
concerning the validity of the procedure leading to the hydrodynamic
description. Indeed, the standard approach retains in the
coarse-grained description only those fields associated with
quantities that are conserved in collisions (such as density and
momentum). There is however good evidence --both numerical and
theoretical-- that in the granular case, a relevant description should
include the kinetic temperature field, defined as the kinetic energy
density (\cite{bclibro2001,d07b} and references therein), which is
therefore not associated to a conserved quantity.

Our purpose here is to test a hydrodynamic description with suitable
coarse-grained fields, for a model system where not only the kinetic
energy is not conserved during binary encounters, but also the number
of particles and the linear momentum. The ballistic annihilation
model \cite{bnklr,bek,ks,t02,ptd02,llf06} provides a valuable candidate: in 
this model, each particle moves freely
(ballistically) until it meets another particle; such binary encounters
lead to the annihilation of the colliding pair of particles.
In addition, we introduce a parameter $0\leq p\leq 1$ that may be thought
of as a measure of the distance to equilibrium, so that an ensemble of
spherical particles in dimension $d$ undergoing ballistic motion
either annihilate upon contact (with probability $p$) or scatter
elastically (with probability $1-p$).  For the corresponding
probabilistic ballistic annihilation model, the Chapman-Enskog
\cite{chapman60} scheme was applied recently \cite{cdt04}.  The
hydrodynamic equations were derived and explicit formulas for the
transport coefficients obtained. Our goal here is two-fold.  First, we
would like to shed light on the context and limitations of the
derivation, by obtaining the hydrodynamic description directly from
the linearized Boltzmann equation. Second, we aim at putting to the
test the theoretical framework thereby obtained by careful comparison
with numerical simulations of the annihilation process.  For granular
gas dynamics, the same program is quite complete, although challenges
remain \cite{g05,d07b}. The objective here is to initiate a similar
formulation for the ballistic annihilation model in view of a more
stringent test of the hydrodynamic machinery.

The paper is organized as follows. We start in section \ref{sec:boltz}
with a reminder of results derived in Refs \cite{t02,ptd02}. The
kinetic description adopted is that of the Boltzmann equation, since
it has been shown that for $p=1$ (all collision events leading to
annihilation), the underlying molecular chaos closure provides an
exact description at long times, provided space dimension $d$ is
strictly larger than $1$ \cite{ptd02}. We may assume that the same holds
for an arbitrary but non vanishing value of $p$, since the density is
then still a decreasing function of time. The focus is here on an unforced
system, which is characterized by an algebraic decay with time of the
total density and kinetic energy density (homogeneous decay state)
\cite{t02,ptd02}. More precisely, we are interested in small
perturbations around this state, so that the Boltzmann equation will
be subsequently linearized. After having identified the operator that
generates the dynamics of fluctuations, attention will be paid in
section \ref{sec:solboltz} to its spectral properties. This will
provide the basis for finding in section \ref{sec:linhyd} the
evolution equations for the hydrodynamic fields (i.e. those chosen for
the coarse-grained description) and for obtaining explicit formulas
for the transport coefficients.  Finally, our predictions will be
confronted in section \ref{sec:md} against extensive Molecular
Dynamics simulations.  Such a comparison is an essential step in
testing the foundations of the hydrodynamic treatment.

\section{The Boltzmann Equation approach to the Homogeneous Decay State}
\label{sec:boltz}
\subsection{Non-linear description}
The Boltzmann equation describes the time evolution of the 
one particle distribution function $f(\mathbf{r},\mathbf{v}_1,t)$. 
For a system of smooth hard disks or spheres of mass
$m$ and diameter $\sigma$, which annihilate with probability $p$ or collide
elastically with probability $1-p$, it has the form  
\begin{equation}\label{b_e}
\left(\frac{\partial}{\partial
t}+\mathbf{v}_1\cdot\nabla\right)f(\mathbf{r},\mathbf{v}_1,t)=pJ_a[f|f]
+(1-p)J_c[f|f], 
\end{equation}
where the annihilation operator $J_a$ is defined by \cite{ptd02}
\begin{equation}
J_a[f|g]=-\sigma^{d-1}\int\!\!
d\mathbf{v}_2\!\!\int\!\! d\sigmagorro\Theta(\mathbf{v}_{12}\cdot\sigmagorro)
(\mathbf{v}_{12}\cdot\sigmagorro)f(\mathbf{r},\mathbf{v}_1,t)
g(\mathbf{r},\mathbf{v}_2,t).
\end{equation}
The elastic collision operator $J_c$ reads \cite{br04,l81}
\begin{equation}
J_c[f|g]=\sigma^{d-1}\int\!\!
d\mathbf{v}_2\!\!\int\!\! d\sigmagorro\Theta(\mathbf{v}_{12}\cdot\sigmagorro)
(\mathbf{v}_{12}\cdot\sigmagorro)(b_{\sigma}^{-1}-1) 
f(\mathbf{r},\mathbf{v}_1,t)g(\mathbf{r},\mathbf{v}_2,t),    
\end{equation}
with $\mathbf{v}_{12}=\mathbf{v}_1-\mathbf{v}_2$, $\Theta$ the Heaviside step
function, $\sigmagorro$ a unit vector joining the centers of the two particles
at contact and $b_{\sigma}^{-1}$ an operator replacing all the velocities
$\mathbf{v}_1$ and $\mathbf{v}_2$ appearing in its argument by their
precollisional values $\mathbf{v}_1^{*}$ and $\mathbf{v}_2^{*}$, given by 
\begin{eqnarray}\label{op_bmenos1}
b_{\sigma}^{-1}\mathbf{v}_1=\mathbf{v}_1^{*}&=&
\mathbf{v}_1-(\mathbf{v}_{12}\cdot\sigmagorro)\sigmagorro,\\ 
b_{\sigma}^{-1}\mathbf{v}_2=\mathbf{v}_2^{*}&=&
\mathbf{v}_2+(\mathbf{v}_{12}\cdot\sigmagorro)\sigmagorro.
\end{eqnarray}
We assume that the system can be characterized macroscopically by coarse
grained (hydrodynamic-like) fields, that we define as in standard Kinetic Theory 
in terms of the local velocity distribution function $f(\mathbf{r},\mathbf{v},t)$
\begin{eqnarray}
n(\mathbf{r},t)&=&\int\!d\mathbf{v}f(\mathbf{r},\mathbf{v},t),\\
n(\mathbf{r},t)\mathbf{u}(\mathbf{r},t)&=&
\int\!d\mathbf{v}\mathbf{v}f(\mathbf{r},\mathbf{v},t),\\
\frac{d}{2}n(\mathbf{r},t)T(\mathbf{r},t)&=&
\int\!d\mathbf{v}\frac{m}{2}V^2f(\mathbf{r},\mathbf{v},t),
\end{eqnarray}
where $n(\mathbf{r},t)$, $\mathbf{u}(\mathbf{r},t)$, and
$T(\mathbf{r},t)$ are the local number density, velocity, and
temperature, respectively. We have introduced here
$\mathbf{V}=\mathbf{v}-\mathbf{u}$, the velocity of the particle
relative to the local velocity flow. We stress that the temperature
defined has a kinetic meaning only, but lacks a thermodynamic
interpretation. It seems natural to consider these fields, as they
are the usual hydrodynamical
fields of the equilibrium system (with $p=0$).
It is however not obvious at this point that restricting our
coarse-grained description to the above three fields provides a
relevant and consistent framework. 
A major goal of this paper is to
provide strong hints that this is indeed the case. We will show in
particular that closed equations can be obtained for these fields in
the appropriate time and length scales, under reasonable assumptions.

The Boltzmann equation (\ref{b_e}) admits a homogeneous scaling
solution $f_{H}$ in which all the time dependence is embedded in the
hydrodynamic fields, with the further simplification that those fields
are position independent. The existence of this regime could not be
shown rigorously, but, numerically, such a scaling solution quickly
emerges from an arbitrary initial condition \cite{t02,ptd02}. It has
the form \cite{ptd02}
\begin{equation}\label{escalamiento_f}
f_H(\mathbf{v},t)=\frac{n_H(t)}{v_H(t)^d}\chi_H(\mathbf{c}),\qquad\hbox{with }
\mathbf{c}=\frac{\mathbf{v}}{v_H(t)},
\end{equation}
where 
\begin{equation}
v_H(t)=\left[\frac{2T_H(t)}{m}\right]^{1/2}
\end{equation}
is the ``thermal'' (root-mean-square) velocity and
$\chi_H(\mathbf{c})$ is an isotropic function depending only on the
modulus $c=|\mathbf{c}|$ of the rescaled velocity. By taking moments
in the Boltzmann equation and using 
the scaling (\ref{escalamiento_f}), it can be seen that the
homogeneous density and temperature obey the equations \cite{cdt04}
\begin{eqnarray}\label{ec_nH}
\frac{\partial n_H(t)}{\partial t}&=&-p\nu_H(t)\zeta_nn_H(t),\\
\label{ec_TH}
\frac{\partial T_H(t)}{\partial t}&=&-p\nu_H(t)\zeta_TT_H(t),
\end{eqnarray} 
where we have introduced the collision frequency 
of the corresponding hard sphere fluid in equilibrium
(with same temperature and density)
\begin{equation}
\nu_H(t)=\frac{n_H(t) \,T^{1/2}_H(t) \,\sigma^{d-1}}{ m^{1/2}}
\frac{8\pi^{\frac{d-1}{2}}}{(d+2)\Gamma(d/2)}
\label{ec_nuH}
\end{equation}
and the dimensionless decay rates $\zeta_n$ and $\zeta_T$, 
that are functionals of the distribution function
\begin{eqnarray}\label{zeta_n}
p\zeta_n&=&-\frac{\gamma}{2}\int\!\!d\mathbf{c}_1\!\!\int\!\!d\mathbf{c}_2
T(\mathbf{c}_1,\mathbf{c}_2)\chi_H(\mathbf{c}_1)\chi_H(\mathbf{c}_2),\\
\label{zeta_T}
p\zeta_T&=&-\frac{\gamma}{2}\int\!\!d\mathbf{c}_1\!\!\int\!\!d\mathbf{c}_2
\left(\frac{2c_1^2}{d}-1\right)T(\mathbf{c}_1,\mathbf{c}_2)
\chi_H(\mathbf{c}_1)\chi_H(\mathbf{c}_2).
\end{eqnarray}
In these expressions, $\gamma$ is a quantity that does not depend on time,
which reads
\begin{equation}
\gamma=\frac{2n_H(t)v_H(t)\sigma^{d-1}}{\nu_H(t)}=
\frac{(d+2)\sqrt{2}\Gamma(d/2)}{4\pi^{(d-1)/2}},
\end{equation}
and the binary collision operator 
$T(\mathbf{c}_1,\mathbf{c}_2)$, that should not be 
confused with the temperature,
takes the form
\begin{equation}
T(\mathbf{c}_1,\mathbf{c}_2)=\int\!\!d\sigmagorro
\Theta(\mathbf{c}_{12}\cdot\sigmagorro)(\mathbf{c}_{12}\cdot\sigmagorro)
[(1-p)b_{\sigma}^{-1}-1]. 
\end{equation}
Finally, we can write an equation for the scaled distribution function 
$\chi_H(\mathbf{c})$ in terms of the coefficients and operators defined above
\begin{equation}\label{ec_chi}
p\left[(d\zeta_T-2\zeta_n)+\zeta_T\mathbf{c}_1\cdot
\frac{\partial}{\partial\mathbf{c}_1}\right]\chi_H(\mathbf{c}_1)=\gamma\int\!\!
d\mathbf{c}_2T(\mathbf{c}_1,\mathbf{c}_2)\chi_H(\mathbf{c}_1)
\chi_H(\mathbf{c}_2). 
\end{equation}
The operator $b_\sigma^{-1}$ in the last equation is defined again by equation
(\ref{op_bmenos1}), but substituting $(\mathbf{v}_1,\mathbf{v}_2)$ by 
$(\mathbf{c}_1,\mathbf{c}_2)$. 

Although an exact and explicit solution of equation
(\ref{ec_chi}) is not known, its behavior at large and small
velocities has been determined \cite{t02,ptd02}. 
In this work we will use the approximate form of the
distribution function in the so-called first Sonine approximation (an
expansion around a Gaussian functional form, see Appendix \ref{appendixA}), which is
valid for velocities in the thermal region, and all the functionals
of $\chi_H(\mathbf{c})$, that is the decay rates and the transport 
coefficients, will be evaluated in this approximation \cite{cdt04s,t02}.

\subsection{Linearized Boltzmann Equation}
In the remainder, we consider a situation where the system is very close to the
homogeneous decay state, so that we can write
\begin{equation}\label{desarrollof}
f(\mathbf{r},\mathbf{v}_1,t)=f_H(\mathbf{v}_1,t)
+\delta f(\mathbf{r},\mathbf{v}_1,t),\qquad
|\delta f(\mathbf{r},\mathbf{v}_1,t)| \ll f_H(\mathbf{v}_1,t).
\end{equation}
Substitution of equation (\ref{desarrollof}) into the Boltzmann equation
(\ref{b_e}), keeping only linear terms in $\delta f$, yields 
\begin{eqnarray}
&&\left(\frac{\partial}{\partial
t}+\mathbf{v}_1\cdot\nabla\right)\delta f(\mathbf{r},\mathbf{v}_1,t)\nonumber\\
&&=p\left\{J_a[\delta f|f_H]+J_a[f_H|\delta f]\right\}
+(1-p)\left\{J_c[\delta f|f_H]+J_c[f_H|\delta f]\right\}, 
\end{eqnarray}

Given the scaling form of $f$ (Eq. (\ref{escalamiento_f})), 
it is convenient to introduce as well the
scaled deviation of the distribution function, $\delta\chi$, as follows
\begin{equation}
\delta f(\mathbf{r},\mathbf{v}_1,t)=\frac{n_H(t)}{v_H(t)^d}
\delta\chi(\mathbf{r},\mathbf{c}_1,\tau) \ .
\end{equation}
Moreover, Eqs. (\ref{ec_nH}) and (\ref{ec_TH}) suggest to use 
the dimensionless time scale $\tau$ defined by
\begin{equation}\label{escala_tau}
\tau=\frac{1}{2}\int_0^tdt'\nu_H(t'), 
\end{equation}
which counts the number of collision per particle in the time
interval $[0,t]$. Combining Eq. (\ref{ec_nuH}) together with
Eq. (\ref{ec_nH}, \ref{ec_TH}) yields immediately $\nu_H(t) =
(1/\nu_H(0) + p(\zeta_n + \zeta_T/2)t)^{-1}$ and thus 
\begin{equation}
\tau=\frac{1}{p (2\zeta_n + \zeta_T)}\log[1+\nu_H(0)p(\zeta_n+\zeta_T/2)t] \;.
\end{equation}
In this time scale $\tau$ (\ref{escala_tau}), these equations
(\ref{ec_nH}, \ref{ec_TH}) are easily integrated, yielding
\begin{equation}
n_H(\tau) = n_H(0) \exp(-2 p \zeta_n \tau), \qquad
T_H(\tau) = T_H(0) \exp(-2 p \zeta_T \tau),
\end{equation}
and power law behaviors in time $t$, $n_H(t) \propto
t^{-2\zeta_n/(2\zeta_n + \zeta_T)}$ and $T_H(t) \propto
t^{-2\zeta_T/(2\zeta_n + \zeta_T)}$ at large time $t\gg 1$ . 
It proves also convenient to introduce Fourier  
components [with the notation ${h}_{\mathbf{k}}=\int\!\!d\mathbf{r}
\exp^{-i\mathbf{k}\cdot\mathbf{r}} h(\mathbf{r})$] so that
the evolution equation for a general $\mathbf{k}$ component of $\delta\chi$ is,
in the $\tau$ timescale,
\begin{equation}\label{l_b_e}
\frac{\partial}{\partial\tau}\delta{\chi}_{\mathbf{k}}(\mathbf{c}_1,\tau)=
\left[\Lambda(\mathbf{c}_1)-il_H(\tau)\mathbf{k}\cdot\mathbf{c}_1\right]
\delta{\chi}_{\mathbf{k}}(\mathbf{c}_1,\tau).
\end{equation}
In this equation, the time dependent length scale 
$l_H =2v_H(\tau)/\nu_H(\tau)$ is proportional to the 
instantaneous mean free path ($l_H(\tau)\propto n_H^{-1}(\tau)$, see
Eq. (\ref{ec_nuH}))
and the homogeneous scaled Boltzmann linear operator reads
\begin{eqnarray}
\Lambda(\mathbf{c}_1)h(\mathbf{c}_1)=\gamma\int\!\!d\mathbf{c}_2
T(\mathbf{c}_1,\mathbf{c}_2)(1+{\cal P}_{12})
\chi_H(\mathbf{c}_1)h(\mathbf{c}_2)
\nonumber\\
+p(2\zeta_n-d\zeta_T)h(\mathbf{c}_1)-p\zeta_T\mathbf{c}_1\cdot
\frac{\partial}{\partial\mathbf{c}_1}h(\mathbf{c}_1).
\end{eqnarray}
In this expression, the permutation operator ${\cal P}_{12}$ interchanges 
the labels of particles $1$ and $2$ and subsequently allows for
more compact notations.
In the present representation, all the time dependence due to the 
reference state is absorbed in the mean free path, obtained from
$l_H(\tau)\propto n_H^{-1}(\tau)$ as
\begin{equation}\label{sol_lH}
l_H(\tau)=l_H(0)\exp(2p\zeta_n\tau) ,
\end{equation}
which, as expected, is an increasing function of time.

\subsection{Linearized Hydrodynamic Equations around the homogeneous decay 
state}
Let us define the relative deviations of the hydrodynamic fields from their 
homogeneous values by
\begin{eqnarray}
\rho(\mathbf{r},\tau)\equiv\frac{\delta n(\mathbf{r},\tau)}{n_H(\tau)}&=&
\int\!\!d\mathbf{c}\delta\chi(\mathbf{r}, \mathbf{c}, \tau),\\
\mathbf{w}(\mathbf{r}, \tau)\equiv
\frac{\delta\mathbf{u}(\mathbf{r},\tau)}{v_H(\tau)}&=&
\int\!\!d\mathbf{c}\mathbf{c}\delta\chi(\mathbf{r}, \mathbf{c}, \tau),\\
\theta(\mathbf{r},\tau)\equiv\frac{\delta T(\mathbf{r},\tau)}{T_H(\tau)}&=&
\int\!\!d\mathbf{c}\left(\frac{2c^2}{d}-1\right)
\delta\chi(\mathbf{r},\mathbf{c},\tau),
\end{eqnarray}
where $\delta y(\mathbf{r},\tau)\equiv y(\mathbf{r},\tau)-y_H(t)$
denotes the deviation of a local macroscopic variable, $
y(\mathbf{r},\tau)$, from its homogeneous decay state value,
$y_H(t)$. Taking velocity moments in the Boltzmann equation
(\ref{l_b_e}), we obtain the linearized balance equation for the
$\mathbf{k}$ components of the hydrodynamic fields
\begin{eqnarray}
\label{ec_l_n}&&\left(\frac{\partial}{\partial\tau}-2p\zeta_n\right)
{\rho}_{\mathbf{k}}
+il_H(\tau)\mathbf{k}\cdot{\mathbf{w}}_{\mathbf{k}}
-p\ \delta\zeta_n[\delta\chi_{\mathbf{k}}]=0,\\
\label{ec_l_u}&&\left[\frac{\partial}{\partial\tau}-p(2\zeta_n+\zeta_T)\right]
{\mathbf{w}}_{\mathbf{k}}\nonumber\\
&&\qquad+\frac{i}{2}l_H(\tau)\mathbf{k}({\rho}_{\mathbf{k}}
+{\theta}_{\mathbf{k}})
+ il_H(\tau)\mathbf{k}
\cdot\boldsymbol\Pi[\delta\chi_{\mathbf{k}}]
-p\ \delta\boldsymbol\zeta_u[\delta\chi_{\mathbf{k}}]=0, \\
\label{ec_l_T}&&\left[\frac{\partial}{\partial\tau}-2p(\zeta_n+\zeta_T)\right]
{\theta}_{\mathbf{k}}\nonumber\\
&&\qquad-2p\zeta_T{\rho}_{\mathbf{k}}+i\frac{2}{d}l_H(\tau)\mathbf{k}\cdot
({\mathbf{w}}_{\mathbf{k}}+\boldsymbol{\phi}[\delta\chi_{\mathbf{k}}])
-p\ \delta\zeta_T[\delta\chi_{\mathbf{k}}]=0.
\end{eqnarray}
Here, we have introduced the traceless pressure tensor and the heat flux as
\begin{eqnarray}\label{def_Pi}
\boldsymbol\Pi[\delta\chi_{\mathbf{k}}]&=&\int\!\!d\mathbf{c}\boldsymbol
\Delta(\mathbf{c})
\delta{\chi}_{\mathbf{k}}(\mathbf{c},\tau),\\
\label{def_phi}\boldsymbol{\phi}[\delta\chi_{\mathbf{k}}]&=&\int\!\!d\mathbf{c}
\boldsymbol{\Sigma}(\mathbf{c})\delta{\chi}_{\mathbf{k}}(\mathbf{c},\tau),
\end{eqnarray}
where $\boldsymbol\Delta$ and $\boldsymbol{\Sigma}$ are defined as
\begin{eqnarray}
\Delta_{ij}(\mathbf{c})&=&c_ic_j-\frac{c^2}{d}\delta_{ij},\\
\boldsymbol\Sigma(\mathbf{c})&=&\left(c^2-\frac{d+2}{2}\right)\mathbf{c},
\end{eqnarray}
and the functionals
\begin{eqnarray}\label{def_zeta_n}
p\ \delta\zeta_n[\delta\chi]=\gamma\int\!\!d\mathbf{c}_1\!\!\int\!\!
d\mathbf{c}_2T(\mathbf{c}_1,\mathbf{c}_2)(1+{\cal P}_{12})\chi_H(\mathbf{c}_1)
\delta{\chi}_{\mathbf{k}}(\mathbf{c}_2,\tau),\\
\label{def_zeta_u}
p\ \delta\boldsymbol\zeta_u[\delta\chi]=\gamma\int\!\!d\mathbf{c}_1\!\!\int\!\!
d\mathbf{c}_2\mathbf{c}_1T(\mathbf{c}_1,\mathbf{c}_2)
(1+{\cal P}_{12})\chi_H(\mathbf{c}_1)
\delta{\chi}_{\mathbf{k}}(\mathbf{c}_2,\tau),\\
\label{def_zeta_T}
p\ \delta\zeta_T[\delta\chi]=\gamma\int\!\!d\mathbf{c}_1\!\!\int\!\!
d\mathbf{c}_2\left(\frac{2c_1^2}{d}-1\right)
T(\mathbf{c}_1,\mathbf{c}_2)(1+{\cal P}_{12})\chi_H(\mathbf{c}_1)
\delta{\chi}_{\mathbf{k}}(\mathbf{c}_2,\tau).  
\end{eqnarray}
The previous analysis  therefore amounts to  obtaining  a set of
complicated equations expressing the evolution of the hydrodynamic
fields as a function of the rescaled homogeneous distribution function
$\chi_H$ and the perturbation $\delta{\chi}$. In order to obtain a
closed set of equations for the hydrodynamic fields (\ref{ec_l_n}),
(\ref{ec_l_u}), (\ref{ec_l_T}), we need therefore to express the
functionals $\boldsymbol\Pi$, $\boldsymbol\phi$, $\delta\zeta_n$,
$\delta\boldsymbol\zeta_u$ and $\delta\zeta_T$, in terms of the
hydrodynamic fields themselves. We will see in the next section that,
as long as we can treat $l_H(\tau)\mathbf{k}$ as a small parameter and 
if the
linear Boltzmann operator has some specific properties, it is possible
to carry out this program and to 
close the linear hydrodynamic equations. However, since the
mean free path $l_H(\tau)$ increases with time (Eq. (\ref{sol_lH})), 
the requirement of a small $l_H(\tau)\mathbf{k}$ is necessarily
limited to a time window depending on both
$\mathbf{k}$ and the probability of annihilation $p$.  An upper bound
for this window is provided by the time when the mean free path
becomes of the order of the system size.

\section{Solution of the Linearized Boltzmann Equation}
\label{sec:solboltz}

In this Section we explore the solutions to the linearized Boltzmann
equation (\ref{l_b_e}) and establish some properties of the
homogeneous linear Boltzmann operator that will be essential for the
coarse-grained description.  From the expression of the linearized
Boltzmann equation, we can identify the operator $\Lambda -
i\mathbf{k}\cdot\mathbf{c}l_H(\tau)$ as the ``generator of the
dynamic'' of $\delta\chi_{\mathbf{k}}$. As we are interested in the
solutions of this equation in the hydrodynamic regime (large enough
scales), it is convenient to study first the eigenvalue problem of the
homogeneous linear Boltzmann operator. The inhomogeneous term will be
treated perturbatively later on.

\subsection{Hydrodynamic Eigenfunctions of $\Lambda$}
Let us consider the eigenvalue problem of the homogeneous linear Boltzmann 
operator $\Lambda$
\begin{equation}\label{problema_autov}
\Lambda(\mathbf{c})\xi_{\beta}(\mathbf{c})=\lambda_{\beta}
\xi_{\beta}(\mathbf{c}).
\end{equation}
Finding all the solutions of this equation is an insurmountable task. 
Nevertheless, it is possible to obtain some
particular solutions, which will turn out to be the relevant ones in the
hydrodynamic regime. The problem will be posed in a Hilbert space of functions 
of $\mathbf{c}$ with scalar product given by 
\begin{equation}\label{scalar_product}
\langle
g|h\rangle=\int\!\!d\mathbf{c}\chi_H^{-1}(\mathbf{c})g^*(\mathbf{c})
h(\mathbf{c}), 
\end{equation}
where $g^*$ denotes the complex conjugate of $g$. 

Of particular interest here are the eigenfunctions and eigenvalues associated
with linear hydrodynamics. Following \cite{bdrlibro2003,bd05}, we use the fact
that the homogeneous decay state is parameterized by the hydrodynamic fields
$n_H$, $T_H$ and $\mathbf{u}_H$. Writing the Boltzmann equation satisfied by
$\chi_H$ and differentiating it with respect to these fields allows then to
obtain three exact relations from which one can extract eigenfunctions of the
linearized Boltzmann collision operator. 
In Appendix \ref{appendixB}, we show that the functions
\begin{eqnarray}\label{gus1}
\xi_1(\mathbf{c})&=&\chi_H(\mathbf{c})+\frac{\partial}{\partial\mathbf{c}}\cdot
\left[\mathbf{c}\chi_H(\mathbf{c})\right],\\\label{gus2}
\xi_2(\mathbf{c})&=&z\chi_H(\mathbf{c})
-\frac{\partial}{\partial\mathbf{c}}\cdot
\left[\mathbf{c}\chi_H(\mathbf{c})\right],\\\label{gus3}
\boldsymbol{\xi}_3(\mathbf{c})&=&-\frac{\partial}{\partial\mathbf{c}}
\chi_H(\mathbf{c}), 
\end{eqnarray}
with $z=2\zeta_n/\zeta_T$,
are solutions of Eq. (\ref{problema_autov}), with
eigenvalues
\begin{equation}
\lambda_1=0, \qquad\lambda_2=-p(\zeta_T+2\zeta_n), \qquad\lambda_3=p\zeta_T, 
\end{equation}
respectively,  $\lambda_3$ being $d$-fold degenerate.
Although we cannot prove in general that these eigenvalues are
indeed the hydrodynamic ones (i.e the upper part of the spectrum), 
we will assume that this is the case ;
the self-consistency of the approach and comparison
with numerical simulations will validate this assumption.
Interestingly, in the particular case of Maxwell molecules
where the full spectrum of $\Lambda$ may be computed exactly
(see Appendix \ref{appendixC}),
it appears that the above ``hydrodynamic'' modes dominate at 
long times, provided that $p<1/4$. For larger values of $p$, 
the ``kinetic'' mode with largest eigenvalue decays slower than 
one of the three ``hydrodynamic'' modes.

As a consequence of the non-hermitian character of the operator $\Lambda$, the
functions $\{\xi_{\beta}\}_{\beta=1,\ldots ,3}$ are not orthogonal with
respect to the scalar product defined in (\ref{scalar_product}).  
They are
nevertheless independent and, in order to define the projection onto
the subspace spanned by these functions, 
it is necessary to introduce a set of functions
$\{\bar{\xi}_{\beta}\}_{\beta=1,\ldots ,3}$ verifying the biorthonormality
condition
\begin{equation} 
\langle\bar{\xi}_{\beta}|\xi_{\beta'}\rangle=\delta_{\beta,\beta'}. 
\end{equation}
Although the set $\{\bar{\xi}_{\beta}\}_{\beta=1,\ldots ,3}$
is not unique, a convenient choice is given by
\begin{eqnarray}\label{gusbarra1}
\bar{\xi}_1(\mathbf{c})&=&\left[\frac{2+z}{2(1+z)}-\frac{z}{1+z}
\frac{c^2}{d}\right]\chi_H(\mathbf{c}),\\\label{gusbarra2}
\bar{\xi}_2(\mathbf{c})&=&\left[\frac{1}{2(1+z)}+\frac{1}{1+z}
\frac{c^2}{d}\right]\chi_H(\mathbf{c}),\\\label{gusbarra3}
\bar{\boldsymbol{\xi}}_3(\mathbf{c})&=&\mathbf{c}\chi_H(\mathbf{c}). 
\end{eqnarray}
Indeed, the functions 
$\{\bar{\xi}_{\beta}\}_{\beta=1,\ldots ,3}$ have to
be linear combinations of  $\chi_H(c)$,
$\mathbf{c}\chi_H(c)$ and $c^2\chi_H(c)$ to ensure that projection
of $\delta \chi_\mathbf{k}$ onto the $\{\bar{\xi}_{\beta}\}$
yields the coarse-grained fields $\rho_\mathbf{k}$,
$\theta_{\mathbf{k}}$ and $\mathbf{w}_{\mathbf{k}}$, or combinations 
thereof. The functions 
$\{\bar{\xi}_{\beta}\}_{\beta=1,\ldots ,3}$ span a dual subspace of that 
spanned by the eigenfunctions and for any linear combination of the
hydrodynamic modes  
\begin{equation}
g(\mathbf{c})=\sum_{\beta=1}^3a_{\beta}\xi_{\beta}(\mathbf{c}), 
\end{equation}
the coefficients $a_{\beta}$ are given by 
\begin{equation}
a_{\beta}=\langle\bar{\xi}_{\beta}|g\rangle=\int\!\!d\mathbf{c}
\chi_H^{-1}(\mathbf{c})\bar{\xi}_{\beta}(\mathbf{c})g(\mathbf{c}). 
\end{equation}
In particular, the projection of 
the distribution function $\delta\chi_\mathbf{k}$ on the subspace spanned by
the functions $\xi_{\beta}$ is given by the coefficients
\begin{equation}
\{\langle\bar{\xi}_{\beta}|\delta\chi_\mathbf{k}\rangle\} = 
\left\{\frac{1}{1+z}\rho_\mathbf{k}
-\frac{z}{2(1+z)}\theta_\mathbf{k},
\frac{1}{1+z}\rho_\mathbf{k}+\frac{1}{2(1+z)}\theta_\mathbf{k},
\mathbf{w}_\mathbf{k}\right\} .
\end{equation}
Notably, these coefficients are simply linear combinations of the
hydrodynamic fields linearized around the homogeneous decay state.

\subsection{Projection of the Linearized Boltzmann Equation on the 
Hydrodynamic  Subspace }
In this section, we study the linearized Boltzmann equation on the
hydrodynamic subspace. Let us define the projectors
\begin{equation}\label{def_P}
Ph(\mathbf{c})=\sum_{\beta=1}^3\langle\bar{\xi}_{\beta}|h\rangle
\xi_{\beta}(\mathbf{c}),
\end{equation}
and
\begin{equation}
P_{\perp}=1-P.
\end{equation}
so that any function can be decomposed as
\begin{equation}
h(\mathbf{c})=Ph(\mathbf{c})+P_{\perp}h(\mathbf{c}).
\end{equation}
In the definition (\ref{def_P}) we are considering the functions 
(\ref{gus1})-(\ref{gus3}) and (\ref{gusbarra1})-(\ref{gusbarra3}) defined
above.

Let us now consider the function $\delta\chi_{\mathbf{k}}$. If we apply the
projectors $P$ and $P_\perp$ to  equation (\ref{l_b_e}), we obtain the
following relations 
\begin{eqnarray}\label{ec_Pdeltachi}
\left[\frac{\partial}{\partial\tau}-P(\Lambda-il_H\mathbf{k}\cdot\mathbf{c})P
\right]P\delta\chi_\mathbf{k}=-Pil_H\mathbf{k}\cdot\mathbf{c}P_\perp
\delta\chi_\mathbf{k}+P\Lambda P_\perp\delta\chi_\mathbf{k},\\
\label{ec_Pdeltachiperp}
\left[\frac{\partial}{\partial\tau}-P_\perp(\Lambda-il_H\mathbf{k}
\cdot\mathbf{c})P_\perp\right]P_\perp\delta\chi_\mathbf{k}
=-P_\perp il_H\mathbf{k}\cdot\mathbf{c}P\delta\chi_\mathbf{k},
\end{eqnarray}
where we have used that 
\begin{equation}
P_\perp\Lambda P=0 \ ,
\end{equation}
which is obtained straightforwardly since $\xi_\beta$ are
right-eigenfunctions of $\Lambda$. We note however that the
$\{\bar{\xi}_{\beta}\}_{\beta=1,\ldots ,3}$ are not
left-eigenfunctions of $\Lambda$, so that $P\Lambda P_\perp\ne 0$.
This also means that $P$ and $\Lambda$ do not commute.

Equations (\ref{ec_Pdeltachi}) and (\ref{ec_Pdeltachiperp}) for
the functions $P\delta\chi_\mathbf{k}$ and  $P_\perp\delta\chi_\mathbf{k}$ are
coupled. Nevertheless, we shall see that, under certain conditions, we 
can decouple the equation for  $P\delta\chi_\mathbf{k}$ in the
long time limit. If we solve formally equation (\ref{ec_Pdeltachiperp}), we 
obtain
\begin{equation}\label{ecP_ortodeltachi}
P_\perp\delta\chi_\mathbf{k}(\mathbf{c},\tau)=
G_0(\tau)P_\perp\delta\chi_\mathbf{k}(\mathbf{c},0)
-\int_0^{\tau}\!\!d\tau'G_{\tau'}(\tau-\tau')P_\perp
il_H(\tau')\mathbf{k}\cdot\mathbf{c}P\delta\chi_\mathbf{k}(\mathbf{c},\tau'),
\end{equation}
where we have introduced the operator $G_{\tau'}(\tau-\tau')$ defined 
from
\begin{equation}
\frac{d}{d\tau}G_{\tau'}(\tau-\tau')=
P_\perp\left[\Lambda(\mathbf{c})-il_H(\tau)\mathbf{k}
\cdot\mathbf{c}\right]P_\perp G_{\tau'}(\tau-\tau'), \qquad 
G_{\tau'}(0)=1.
\end{equation}
If the hydrodynamic eigenvalues of the operator $\Lambda$ are separated enough 
from the rest of the spectrum, the first term on the right hand side of 
(\ref{ecP_ortodeltachi}) decays with the ``non hydrodynamic'' 
modes,  faster than the second one. We can then write
\begin{equation}\label{ec_Pdeltachiperp_2}
P_\perp\delta\chi_\mathbf{k}(\mathbf{c},\tau)\approx
-\int_0^{\tau}\!\!d\tau'G_{\tau'}(\tau-\tau')
P_\perp il_H(\tau-\tau')\mathbf{k}\cdot\mathbf{c}
P\delta\chi_\mathbf{k}(\mathbf{c},\tau-\tau').
\end{equation}
and we see, by substituting equation (\ref{ec_Pdeltachiperp_2}) in 
(\ref{ec_Pdeltachi}), that we obtain an involved but closed equation for
$P\delta\chi_\mathbf{k}$. It is worth pointing out that we have not proved
scale separation, but assumed it in order to derive 
(\ref{ec_Pdeltachiperp_2}). For an explicit discussion of the scale
separation assumption in a similar but somewhat simplified context, 
we refer to Appendix \ref{appendixC},
already alluded to above. 

The set of  hydrodynamic equations
(\ref{ec_l_n})-(\ref{ec_l_T}) have been obtained through 
the projection of the Boltzmann equation
onto the hydrodynamic subspace. It now appears that, in the
hydrodynamic time scale, the use of Eq. (\ref{ec_Pdeltachiperp_2})
will allow us to close these equations by substituting
the distribution function by its decomposition in terms of the
projectors,
$\delta\chi_\mathbf{k}=P\delta\chi_\mathbf{k}+P_\perp\delta\chi_\mathbf{k}$.
This is the aim of the next section.

\section{Linear Hydrodynamic Equations in 
Navier-Stokes order}
\label{sec:linhyd}
In this section we shall use the decomposition of
$\delta\chi_{\mathbf{k}}$ into its hydrodynamic part,
$P\delta\chi_{\mathbf{k}}$, and non-hydrodynamic part,
$P_\perp\delta\chi_{\mathbf{k}}$, to close the linear hydrodynamic
equations (\ref{ec_l_n})-(\ref{ec_l_T}). We shall do so in
Navier-Stokes order, that is, in the long time limit and in second
order in the gradients (order $k^2$).

Let us first introduce 
$P\delta\chi_{\mathbf{k}}$ in the linear pressure tensor 
and in the heat flux vector. Here the calculation is straightforward and we 
obtain
\begin{equation}
\boldsymbol{\Pi}[P\delta\chi_{\mathbf{k}}]=\mathbf{0},\qquad
\boldsymbol{\phi}[P\delta\chi_{\mathbf{k}}]=\mathbf{0},
\end{equation}
because the functions $\chi_H(\mathbf{c})\boldsymbol{\Delta}(\mathbf{c})$ and
$\chi_H(\mathbf{c})\boldsymbol{\Sigma}(\mathbf{c})$ are orthogonal to the 
subspace spanned by the hydrodynamic eigenfunctions 
$\{\xi_\beta(\mathbf{c})\}_{\beta=1,...3}$.
Turning our attention   
to the other functionals 
$\delta\zeta_n$, $\delta\boldsymbol\zeta_u$ and $\delta\zeta_T$, the
calculations become somewhat lengthy, and we show the details in
the Appendix \ref{appendixD}. We obtain
\begin{eqnarray}
\delta\zeta_n[P\delta\chi_{\mathbf{k}}]&=&-4\zeta_n\rho_\mathbf{k}
-\zeta_n\theta_\mathbf{k},\\
\delta\boldsymbol{\zeta}_u[P\delta\chi_{\mathbf{k}}]&=&
-2\zeta_n\mathbf{w}_\mathbf{k},\\
\delta\zeta_T[P\delta\chi_{\mathbf{k}}]&=&-4\zeta_T\rho_\mathbf{k}
-(3\zeta_T+2\zeta_n)\theta_\mathbf{k}.
\label{eq:resuapp}
\end{eqnarray}
The negative signs occurring on the right hand side of these relations
account for the fact that a fluctuation with a local enhanced density
will induce an increased collision rate, hence a faster density decay.
The same remark holds for temperature or local velocity flow fluctuations.

We now have to calculate the contribution of
$P_\perp\delta\chi_{\mathbf{k}}$ to the same functionals, to second
order in $k$. This requires the knowledge of
$P_\perp\delta\chi_{\mathbf{k}}$ to first order in $k$ since the heat
flux and pressure tensor enter the balance equations
(\ref{ec_l_n})-(\ref{ec_l_T}) through their gradients and are already
weighted by a factor $k$. However, it should be noted that for
consistency, the decay rates should be computed to second order in the
gradients (see Eqs. (\ref{ec_l_n})-(\ref{ec_l_T})).  We shall
nevertheless restrict to first order, henceforth neglecting the
various terms of order two that symmetry allows (such as $\nabla^2 n$
and $\nabla^2 T$ for $\delta\zeta_n$ and $\delta\zeta_T$, or as
$\nabla^2 \mathbf{u}$ for $\delta\boldsymbol{\zeta}_u$).  We will
further comment this approximation below.  To leading order, we have
that $G_{\tau-\tau'}(\tau')\approx e^{P_\perp\Lambda P_\perp\tau'}$,
so that we obtain from equation (\ref{ec_Pdeltachiperp_2})
\begin{eqnarray}\label{ec_Pdeltachiperp_3}
&&P_\perp\delta\chi_\mathbf{k}(\mathbf{c},\tau)\approx\nonumber\\ 
&&\approx-\int_0^{\tau}\!\!d\tau'e^{ P_\perp\Lambda P_\perp\tau'}
P_\perp il_H(\tau-\tau')\mathbf{k}\cdot\mathbf{c}
P\delta\chi_\mathbf{k}(\mathbf{c},\tau-\tau')\nonumber\\
&&\approx-l_H(\tau)
\int_0^{\tau}\!\!d\tau'e^{ P_\perp\Lambda P_\perp\tau'}
e^{-2p\zeta_n\tau'}P_\perp i\mathbf{k}\cdot\mathbf{c}
P\delta\chi_\mathbf{k}(\mathbf{c},\tau-\tau'), 
\end{eqnarray}
where we have used that $l_H(\tau)\propto e^{2p\zeta_n\tau}$ 
(Eq. (\ref{sol_lH})). 
We now have to relate 
$P\delta\chi_\mathbf{k}(\tau-\tau')$ to $P\delta\chi_\mathbf{k}(\tau)$, and
to be consistent with the approximation made above, we also have to restrict 
to leading order in $k$. In doing so, Markovian equations for the fields
will be derived.
From Eq. (\ref{ec_Pdeltachi}), we get
\begin{equation}\label{ec_Pdeltachi0}
P\delta\chi_\mathbf{k}(\mathbf{c},\tau-\tau')\approx e^{-P\Lambda P\tau'}
P\delta\chi_\mathbf{k}(\mathbf{c},\tau)=\sum_{\beta=1}^3
e^{-\lambda_{\beta}\tau'}\langle
\bar{\xi}_\beta|\delta\chi_\mathbf{k}(\tau)\rangle\xi_\beta(\mathbf{c}).
\end{equation}
Substituting (\ref{ec_Pdeltachi0}) in (\ref{ec_Pdeltachiperp_3}),
we obtain an equation for $P_\perp\delta\chi_\mathbf{k}$ to first order in $k$
\begin{equation}\label{ec_Pdeltachiperp_4}
P_\perp\delta\chi_\mathbf{k}^{(1)}(\mathbf{c},\tau)=
-l_H(\tau)\sum_{\beta=1}^3\langle
\bar{\xi}_\beta|\delta\chi_\mathbf{k}(\tau)\rangle
\int_0^{\tau}\!\!d\tau'
e^{ P_\perp(\Lambda-2p\zeta_n-\lambda_\beta)P_\perp\tau'}
P_\perp i\mathbf{k}\cdot\mathbf{c}
\xi_\beta(\mathbf{c}), 
\end{equation}
where $P_\perp\delta\chi_\mathbf{k} =
P_\perp\delta\chi_\mathbf{k}^{(1)} + {\cal O}(k^2)$. The pressure
tensor and the heat flux up to first order in the gradients of 
the fields are now obtained by substituting equation (\ref{ec_Pdeltachiperp_4})
into equations (\ref{def_Pi}) and (\ref{def_phi}). Taking into account the
symmetry properties of the system, the resulting expressions can be written in
the form 
\begin{eqnarray}\label{ec_Pi}
\Pi_{ij}[P_\perp\delta\chi_\mathbf{k}^{(1)}]
&=&-il_H(\tau)\tilde{\eta}(\tau)\left[k_jw_{i,\mathbf{k}}
+k_iw_{j,\mathbf{k}}
+\frac{2}{d}\mathbf{k}\cdot\mathbf{w}_\mathbf{k}\delta_{ij}\right],\\
\label{ec_phi}\boldsymbol{\phi}[P_\perp\delta\chi_\mathbf{k}^{(1)}]
&=&-il_H(\tau)\mathbf{k}
\left[\tilde{\kappa}(\tau)\theta_{\mathbf{k}}
+\tilde{\mu}(\tau)\rho_{\mathbf{k}}\right].
\end{eqnarray}
Equation (\ref{ec_Pi}) is the expected Navier-Stokes expression for the
pressure tensor, involving the shear viscosity coefficient $\tilde{\eta}$, but
equation (\ref{ec_phi}) contains, besides the usual Fourier law characterized
by the heat conductivity $\tilde{\kappa}$, an additional contribution
proportional to the density gradient and with an associated transport
coefficient $\tilde{\mu}$. This latter term is analogous to the one appearing
in granular gases \cite{bte05,d07}. 

The expression of the (time dependent) transport coefficients are
\begin{equation}\label{eta}
\tilde{\eta}(\tau)=\int\!\!d\mathbf{c}\Delta_{xy}(\mathbf{c})
F_{3,xy}(\mathbf{c},\tau)
=\frac{1}{d^2+d-2}\sum_{i,j}^d\int\!\!d\mathbf{c}\Delta_{ij}(\mathbf{c})
F_{3,ij}(\mathbf{c},\tau),
\end{equation}
\begin{eqnarray}\label{mu}
\tilde{\mu}(\tau)=\frac{1}{d(1+z)}\int\!\!d\mathbf{c}
\boldsymbol{\Sigma}(\mathbf{c})\left[\mathbf{F}_1(\mathbf{c},\tau)
+\mathbf{F}_2(\mathbf{c},\tau)\right],\\
\label{kappa}
\tilde{\kappa}(\tau)=\frac{1}{2d(1+z)}\int\!\!d\mathbf{c}
\boldsymbol{\Sigma}(\mathbf{c})\left[-z\mathbf{F}_1(\mathbf{c},\tau)
+\mathbf{F}_2(\mathbf{c},\tau)\right], 
\end{eqnarray}
where we have introduced the functions
\begin{eqnarray}
F_{3,ij}(\mathbf{c},\tau)&=&\int_0^\tau\!\!d\tau'
e^{P_\perp(\Lambda-2p\zeta_n-p\zeta_T)\tau'}
P_\perp c_i\xi_{3,j}(\mathbf{c}),
\label{eq:F3}\\
\mathbf{F}_1(\mathbf{c},\tau)&=&\int_0^\tau\!\!d\tau'
e^{P_\perp(\Lambda-2p\zeta_n)\tau'}P_\perp\mathbf{c}\xi_1(\mathbf{c}), \\
\mathbf{F}_2(\mathbf{c},\tau)&=&\int_0^\tau\!\!d\tau'
e^{P_\perp(\Lambda+p\zeta_T)\tau'}P_\perp\mathbf{c}
\xi_2(\mathbf{c}), 
\label{eq:F2}
\end{eqnarray}
and in the second equality of equation (\ref{eta}), we have summed
over all the $i$, $j$, taking into account the symmetry of the
linearized Boltzmann operator.

Similarly, we calculate the deviations of the decay rates to first order in 
the gradients of the fields by substituting equation (\ref{ec_Pdeltachiperp_4})
into equations (\ref{def_zeta_n}), (\ref{def_zeta_u}) and (\ref{def_zeta_T}). 
Taking into account the symmetry properties, we arrive at
\begin{eqnarray}\label{deltazetan}
\delta\zeta_n[P_\perp\delta\chi_\mathbf{k}^{(1)}]&=&0, \\
\label{deltazetau}
\delta\boldsymbol{\zeta}_u[P_\perp\delta\chi_\mathbf{k}^{(1)}]&=&
il_H(\tau)\mathbf{k}\left[\zeta_{u,\rho}(\tau)\rho_\mathbf{k}
+\zeta_{u,\theta}(\tau)\theta_\mathbf{k}\right], \\
\label{deltazetaT}
\delta\zeta_T[P_\perp\delta\chi_\mathbf{k}^{(1)}]&=&0.
\end{eqnarray}
The expression for the coefficients are
\begin{equation}\label{zeta_u_rho}
\zeta_{u,\rho}(\tau)=\frac{\gamma\beta}{d(1+z)}\int\!\!d\mathbf{c}_1\!\!
\int\!\!d\mathbf{c}_2\chi_H(\mathbf{c}_1)c_{12}(\mathbf{c}_1+\mathbf{c}_2)\cdot
\left[\mathbf{F}_1(\mathbf{c}_2,\tau)+\mathbf{F}_2(\mathbf{c}_2,\tau)\right],
\end{equation}
\begin{equation}\label{zeta_u_theta}
\zeta_{u,\theta}(\tau)=\frac{\gamma\beta}{2d(1+z)}\int\!\!d\mathbf{c}_1\!\!
\int\!\!d\mathbf{c}_2\chi_H(\mathbf{c}_1)c_{12}(\mathbf{c}_1+\mathbf{c}_2)\cdot
\left[-z\mathbf{F}_1(\mathbf{c}_2,\tau)+\mathbf{F}_2(\mathbf{c}_2,\tau)\right],
\end{equation}
with $\beta=\pi^{(d-1)/2}/\Gamma[(d+1)/2]$ the $d$-dimensional solid angle.

At this point, it is important to note that the transport coefficients defined
in equations (\ref{eta})-(\ref{kappa}) and
(\ref{zeta_u_rho})-(\ref{zeta_u_theta}) are time-dependent, and this
dependence is governed by the $\mathbf{F}_i$ functions.  The (exponential)
integrands appearing in the definitions of the $\mathbf{F}_i$ decay with the
non-hydrodynamic (kinetic) modes, as a consequence of the action of the
projector $P_\perp$. From our assumption of hydrodynamic versus kinetic scale
separation, all kinetic eigenvalues are smaller than the smallest hydrodynamic
eigenvalue of $\Lambda$, which is $\lambda_2=-p\zeta_T-2p\zeta_n$. This
ensures the convergence of the integrals (\ref{eq:F3})-(\ref{eq:F2}) for $\tau
\to \infty$. In order for the transport coefficients to reach their
$\tau\to \infty$ limit faster than any of the hydrodynamic time scales, we
need moreover the more stringent condition that the fastest kinetic mode is at
least separated by a $p\zeta_T$ gap from $\lambda_2=-p\zeta_T-2p\zeta_n$: under
this condition, the time dependence of the exponential term in the integral
giving $\mathbf{F}_2$ is fast enough so that the transport coefficients, that
depend on the $\mathbf{F}_i$ functions through (\ref{eta})-(\ref{kappa}), can
be considered as constants on the hydrodynamic time scale.  With this proviso
in mind, it is possible to set $\tau\to \infty$ in the integrals
(\ref{eq:F3})-(\ref{eq:F2}) and the time-independent transport coefficients
obtained in this section are then equivalent to those calculated in reference
\cite{cdt04} by the Chapman-Enskog method. We recall in Appendix \ref{appendixA}
their expressions in the first order Sonine approximation.

Finally, if we substitute the expressions derived above for the
fluxes, equations (\ref{ec_Pi})-(\ref{ec_phi}), and the decay rates,
equations (\ref{deltazetan})-(\ref{deltazetaT}), and we take into
account that in the hydrodynamic time scale we can substitute all the
coefficients by their $\tau\rightarrow\infty$ limit, we obtain the
following closed equations for the linear deviation of the
hydrodynamic fields
\begin{eqnarray}
\label{ec_l_n1}
&&\left(\frac{\partial}{\partial\tau}+2p\zeta_n\right)
{\rho}_{\mathbf{k}}
+il_H(\tau)kw_{\mathbf{k}||}
+p\zeta_n\theta_\mathbf{k}=0,\\
\label{ec_l_u1t}
&&\left[\frac{\partial}{\partial\tau}-p\zeta_T 
+l_H^2(\tau)\tilde{\eta}k^2\right]
{\mathbf{w}}_{\mathbf{k}\perp}=0, \\ 
\label{ec_l_u1p}
&&\left[\frac{\partial}{\partial\tau}-p\zeta_T
+\frac{2(d-1)}{d}l_H^2(\tau)\tilde{\eta}k^2\right]
w_{\mathbf{k}||}\nonumber\\
&&\qquad+\frac{i}{2}l_H(\tau)k\left[(1-2p\zeta_{u,\rho}){\rho}_{\mathbf{k}}
+(1-2p\zeta_{u,\theta}){\theta}_{\mathbf{k}}\right]=0, \\
\label{ec_l_T1}
&&\left[\frac{\partial}{\partial\tau}+p\zeta_T
+\frac{2}{d}l_H^2(\tau)\tilde{\kappa}k^2\right]{\theta}_{\mathbf{k}}
\nonumber\\ 
&&\qquad+\left[2p\zeta_T+\frac{2}{d}l_H^2(\tau)\tilde{\mu}k^2\right]
\rho_\mathbf{k}
+i\frac{2}{d}l_H(\tau)kw_{\mathbf{k}||}=0,
\end{eqnarray}
where $w_{\mathbf{k}||}$ and $\mathbf{w}_{\mathbf{k}\perp}$ are the
longitudinal and transversal parts of the velocity vector defined by
\begin{equation}
w_{\mathbf{k}||}=\mathbf{w}_\mathbf{k}\cdot\hat{\mathbf{k}},\quad 
\mathbf{w}_{\mathbf{k}\perp}=\mathbf{w}_\mathbf{k}-w_{\mathbf{k}||}
\hat{\mathbf{k}},
\end{equation}
and $\hat{\mathbf{k}}$ is the unit vector along the direction given by 
$\mathbf{k}$. 

Equation (\ref{ec_l_u1t}) for the shear mode is decoupled from
the other equations and can be readily integrated. If we introduce a
non-dimensional wave number $\tilde{k}=l_H(0)k$, scaled by the mean free path
at the time origin, we obtain the explicit solution
\begin{equation}\label{solucion_wp}
\mathbf{w}_{\mathbf{k}\perp}(\tau)=
\exp\left[p\zeta_T\tau-\frac{\tilde{\eta} \tilde{k}^2}{4p\zeta_n}
\left(e^{4p\zeta_n\tau}-1\right)\right]\mathbf{w}_{\mathbf{k}\perp}(0).
\end{equation}
Interestingly, depending on $\tilde{k}$, the perturbation may initially
increase if $p\zeta_T-\tilde{\eta}\tilde{k}^2 > 0$. For 
long times however, the exponential $e^{4p\zeta_n\tau}$ 
always dominates the linear term $p\zeta_T\tau$ and the perturbation decays.

The other three fields, namely, the density $\rho_\mathbf{k}$, temperature 
$\theta_\mathbf{k}$, and the longitudinal velocity $w_{\mathbf{k} ||}$, obey
the system of coupled linear equations
\begin{equation}\label{sistema}
\frac{\partial}{\partial\tau}
\left(\begin{array}{c}
\rho_\mathbf{k}\\
w_{\mathbf{k} ||}\\
\theta_\mathbf{k}\\
\end{array}\right)
=\mathbf{M}(\tau)\cdot
\left(\begin{array}{c}
\rho_\mathbf{k}\\
w_{\mathbf{k} ||}\\
\theta_\mathbf{k}
\end{array}\right), 
\end{equation}
where the time-dependent matrix is
\begin{eqnarray}
\mathbf{M}(\tau)=
\left(\begin{array}{ccc}
-2p\zeta_n\ & -il_H(\tau)k & -p\zeta_n\\
-\frac{i}{2}l_H(\tau)k(1-2p\zeta_{u,\rho}) &
p\zeta_T-\frac{2(d-1)}{d}l_H^2(\tau)\tilde{\eta}k^2 &
-\frac{i}{2}l_H(\tau)k(1-2p\zeta_{u,\theta})\\
-2p\zeta_T-\frac{2}{d}l_H^2(\tau)\tilde{\mu}k^2 &
-i\frac{2}{d}l_H(\tau)k &
-p\zeta_T-\frac{2}{d}l_H^2(\tau)\tilde{\kappa}k^2
\end{array}\right).\nonumber\\
\label{eq:hydromat}
\end{eqnarray}
We note here that this matrix differs from Eq. (59) of Ref.
\cite{cdt04}, where the analysis amounts to overlooking the time
dependence of the mean free path, so that all entries of the
hydrodynamic matrix exhibit the same time dependence. The different
time dependences present in Eq. (\ref{eq:hydromat}) render the
stability analysis more difficult. For long times however, the
eigenvalues of the matrix $\mathbf{M}(\tau)$ are always negative and
the perturbations {\it a priori} decay.  A {\it caveat} is
nevertheless in order. It is worth pointing out that equations
(\ref{solucion_wp}) and (\ref{sistema}) break down at long
times. Once the function $l_H(\tau)k$
exceeds
unity indeed, the expansion in the gradients we have performed is no longer
valid, and it would be necessary to include terms of higher order in
$k$. In addition, if the perturbation initially increases sufficiently
to leave the linear regime, our description breaks down and it becomes
necessary to consider non linear terms.

Although quite involved, the evolution equations (\ref{sistema}) can be 
numerically integrated, using for instance the transport coefficients
computed in the Sonine approximation in Appendix \ref{appendixA}. This is
what we do in the next section, in order to compare the theoretical
predictions with Molecular Dynamics simulations.

\section{Molecular Dynamics simulations}\label{simulaciones}
\label{sec:md}

To put our theoretical predictions to the test, we have performed Molecular
Dynamics (MD) simulations of a system of $N$ smooth hard disks ($d=2$) which
undergo ballistic flights punctuated by collisional events : at each
collision, the discs annihilate with probability $p$ ; otherwise, they collide
elastically. The particles are localized in a square box of size $L$ with
periodic boundary conditions. An event driven algorithm \cite{allen} has been
used and the initial density has been chosen low enough to be always in the
dilute limit. The parameters for all the MD simulations are $N(0)=10^5$,
reduced density $n(0)\sigma^2=0.05$ where $\sigma$ is the discs' radius and
$0<p\le 1$.  The initial conditions we have considered correspond to small
amplitude perturbations around the homogeneous decay state, to enforce the
validity of the linearized hydrodynamic equations
(\ref{ec_l_n1})-(\ref{ec_l_T1}).

\begin{figure}
\begin{minipage}{0.46\linewidth}
\includegraphics[angle=0,width=0.9\linewidth]
{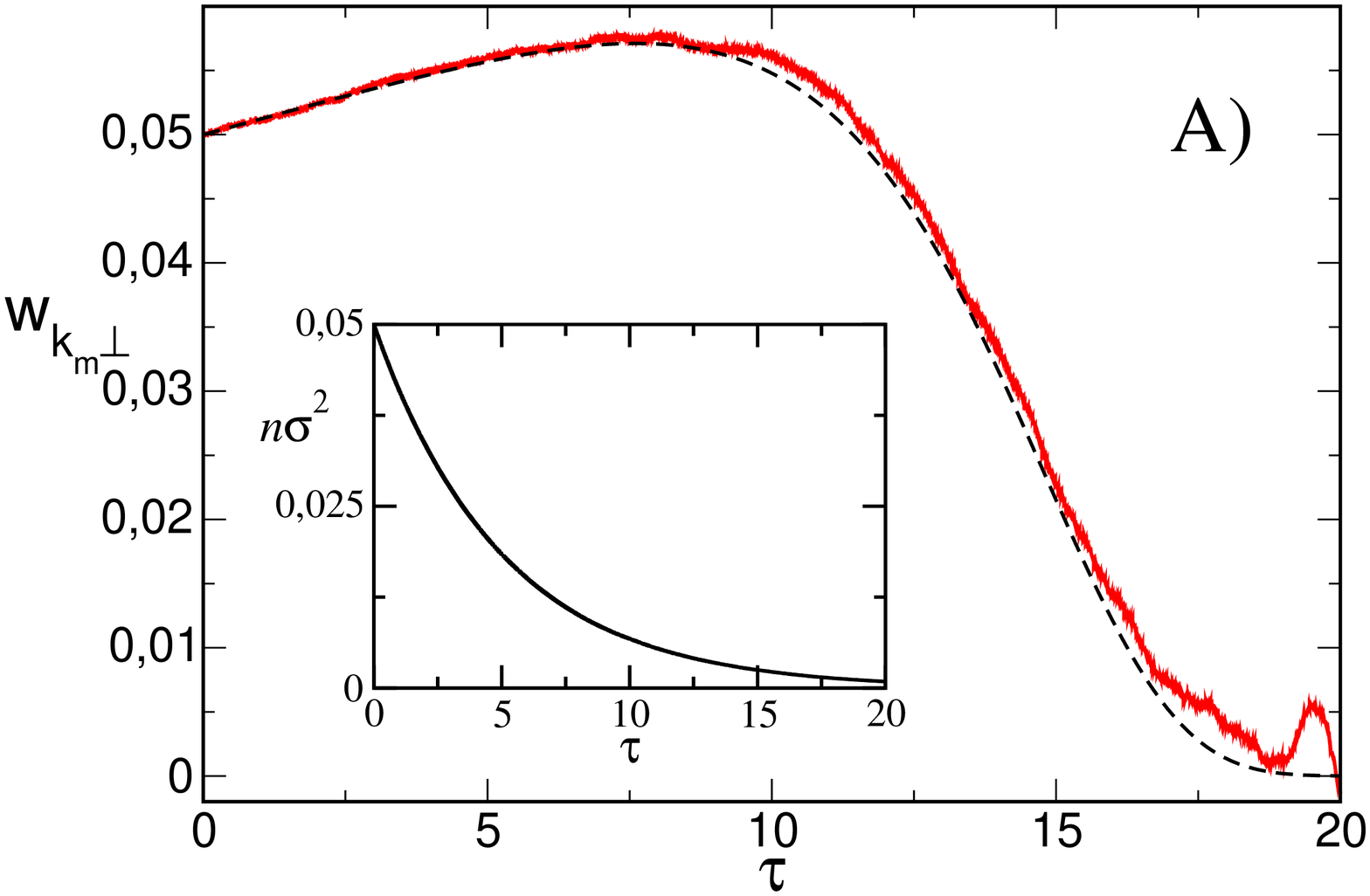}
\end{minipage}\hfill
\begin{minipage}{0.46\linewidth}
\vspace*{0.2cm}
\includegraphics[angle=0,width=0.9\linewidth]
{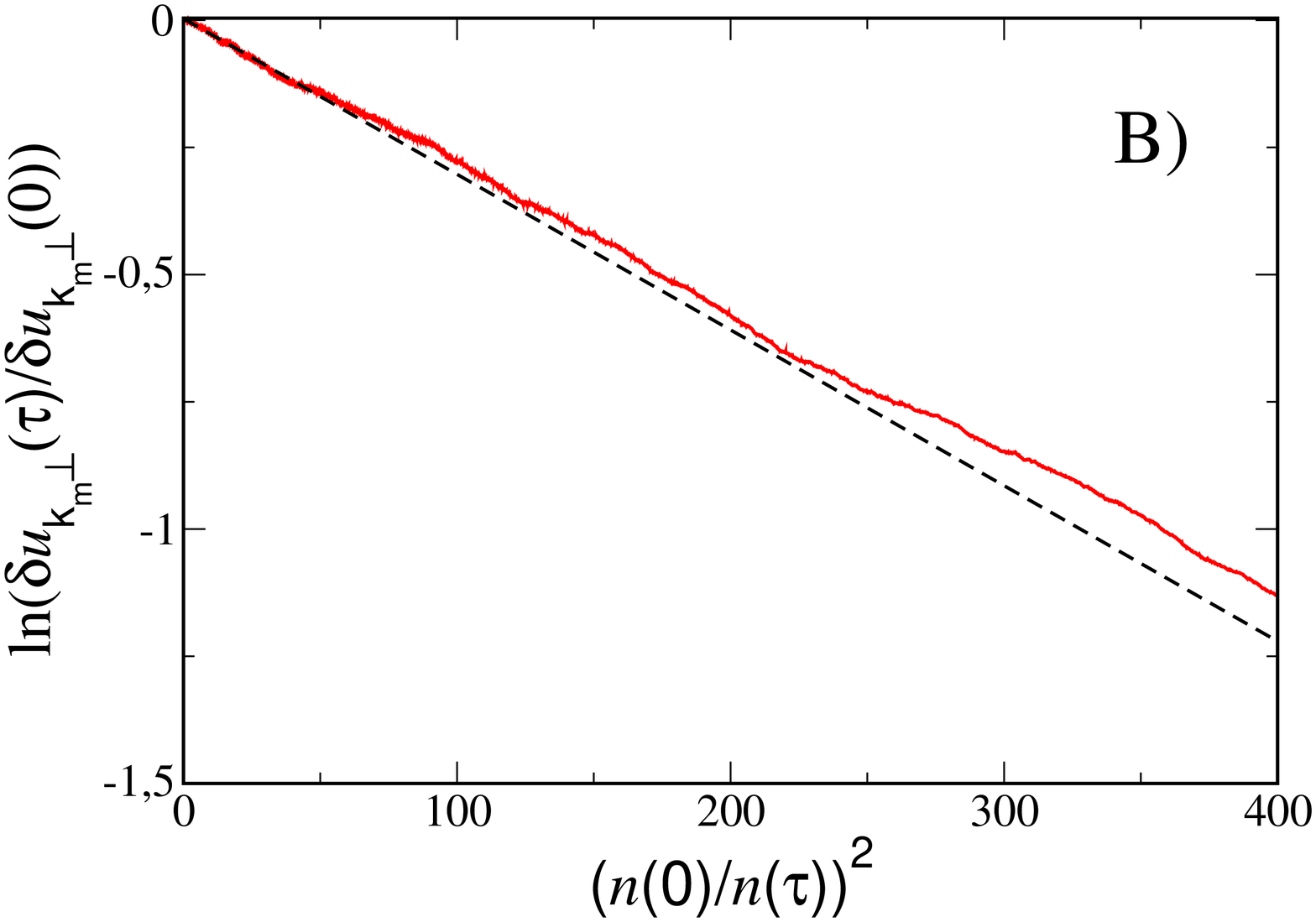}
\end{minipage}
\caption{A) Time evolution of a linear perturbation of the transversal velocity
  for a system with annihilation probability $p=0.1$. The solid lines are the
  Molecular Dynamics results and the dashed lines are the theoretical
  predictions (no adjustable parameter). Note how the theoretical
predictions correctly account for the increase of the perturbation at
short times.
The inset shows the evolution of
density with rescaled time $\tau$, where $\sigma$ is the discs' radius. 
For $\tau=20$, $n\sigma^2 \simeq 9 . 10^{-4}$.
B) Evolution of 
$\delta u_{\mathbf{k}\perp}(\tau)/\delta u_{\mathbf{k}\perp}(0)$ 
as a function of $n^2(0)/n^2(\tau)$. The
dashed line is the exponential decay predicted
by Eq. (\ref{solucion_up}).  $n^2(0)/n^2(\tau)=400$ corresponds to $\tau\simeq
15$.
}\label{vtransversalp0.1}
\end{figure}

\begin{figure}
\begin{minipage}{0.46\linewidth}
\includegraphics[angle=0,width=0.9\linewidth]
{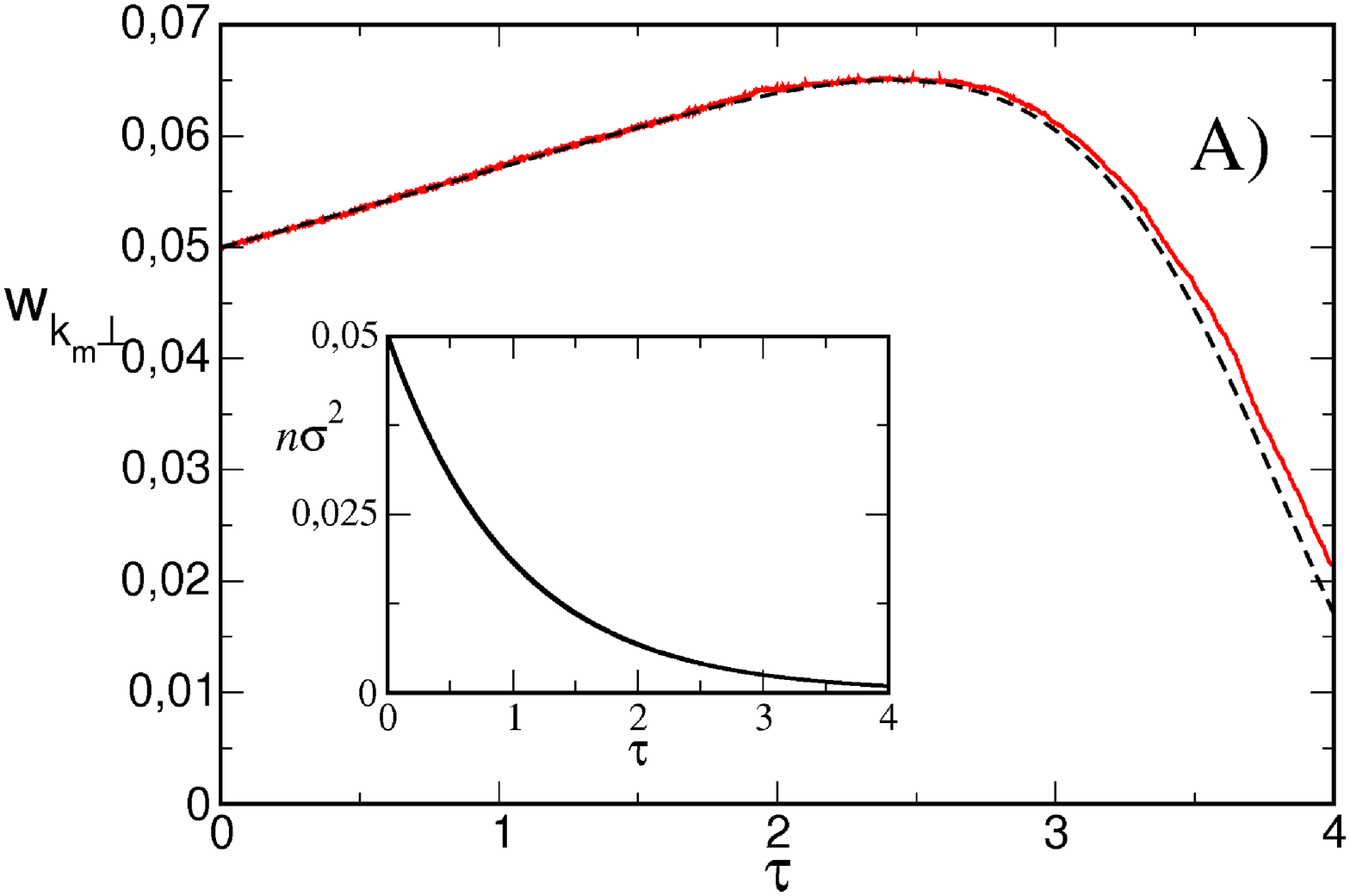}
\end{minipage}\hfill
\begin{minipage}{0.46\linewidth}
\vspace*{0.3cm}
\includegraphics[angle=0,width=0.9\linewidth]
{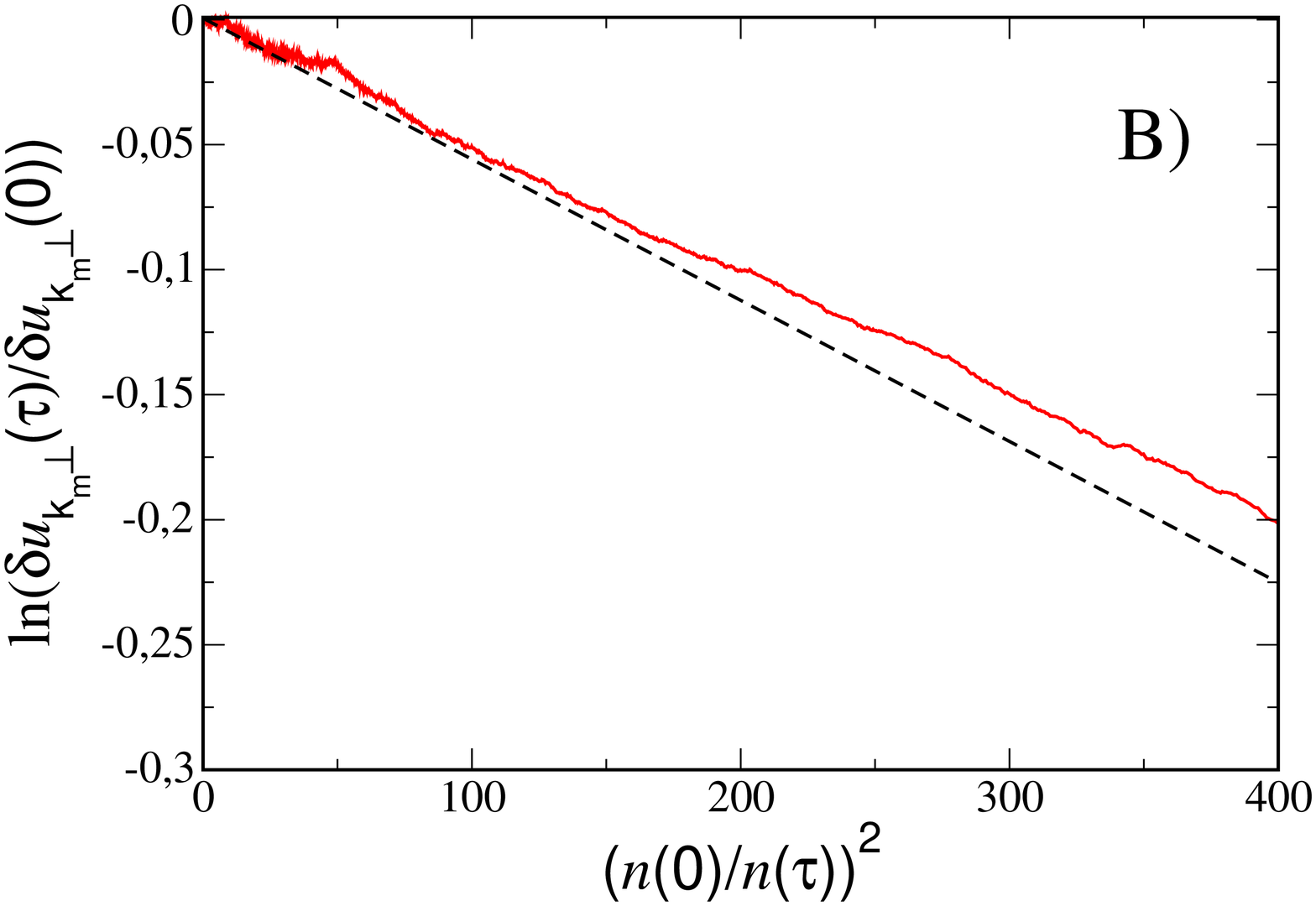}
\end{minipage}
\caption{A) Same as in Fig. \ref{vtransversalp0.1} 
but for a system with $p=0.5$, which --loosely speaking-- may therefore
be considered as being more ``distant'' to equilibrium. The inset shows the 
evolution of density with rescaled time $\tau$. For $\tau=4$,
the rescaled density is $n\sigma^2 \simeq 9 . 10^{-4}$ (while
$n(0)\sigma^2=0.05$). B) Evolution of 
$\delta u_{\mathbf{k}\perp}(\tau)/\delta u_{\mathbf{k}\perp}(0)$ 
as a function of $n^2(0)/n^2(\tau)$.
$n^2(0)/n^2(\tau)=400$ corresponds to $\tau\simeq 3$. }
\label{vtransversalp0.5}
\end{figure}

\begin{figure}
\includegraphics[angle=0,width=0.46\textwidth]{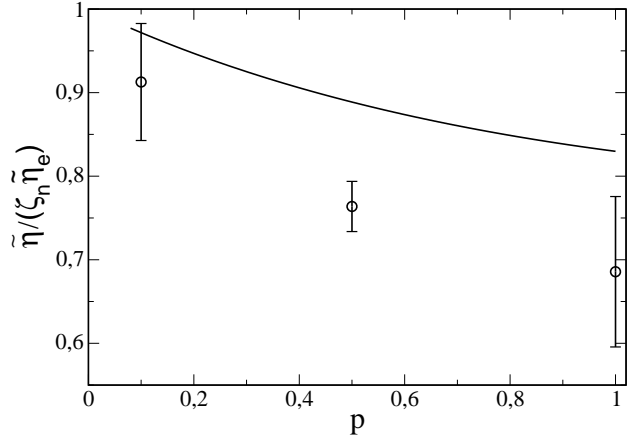}
\caption{Symbols: ratio $\tilde{\eta}/\zeta_n$ (normalized by its $p\to 0$
  value) extracted from the exponential fit of $\delta
  u_{\mathbf{k}\perp}(\tau)$ to Eq. (\ref{solucion_up}).  The solid line is
  the theoretical prediction in the first Sonine approximation.}
\label{etas}
\end{figure}

\subsection{Perturbation of the transversal velocity}

Since Eq. (\ref{ec_l_u1t}) for the shear mode is
decoupled from the other equations, one of
the simplest macroscopic perturbation one
can think of consists in an initial harmonic perturbation of the transversal
component of the velocity field, whose evolution is given by Eq. 
(\ref{solucion_wp}). We shall consider a small perturbation in real
space of the form
\begin{equation}
u_x(\mathbf{r},0)=A\sin(k_{m}y),
\end{equation}
with $A=10^{-1} v_H(0)$ and $k_m=2\pi/L$, where $L$ is the linear size
of the system. The reason for choosing the smallest possible value of
$k$ compatible with the boundary conditions is twofold. First, the
corresponding mode is the most unstable at short times (see
Eq. (\ref{solucion_wp})).  For the parameters of the simulations, we
indeed probe the region where
$p\zeta_T>\tilde{\eta}\tilde{k}^2$, so that 
the hydrodynamic equation (\ref{solucion_wp}) predicts an 
initial {\em increase} of the
perturbation. Second, the low $k$ regime is
that where our large scale predictions are most likely to be relevant.

Figure \ref{vtransversalp0.1} displays the evolution of
$w_{k_m\perp}$ as a function of the number of collisions
per particle $\tau$ for a system with $p=0.1$,
averaging data over $50$ different trajectories. The solid line is the
simulation result and the dashed line is the theoretical prediction,
Eq. (\ref{solucion_wp}), where the shear viscosity and the decay
rates have been computed using the standard tools of kinetic theory
(here, the first Sonine approximation \cite{cdt04}, see Appendix
\ref{appendixA}). An excellent agreement is obtained {\em without
any adjustable parameter}, including the 
predicted increase of the perturbation at short times, 
and as also observed for a larger annihilation probability
$p=0.5$ (see Fig. \ref{vtransversalp0.5}, in which the MD data are
obtained by an average over $150$ runs).

Recalling that $n_H(\tau)/n_H(0) = \exp(-2 p \zeta_n
\tau)$ and that $v_H(\tau) = v_H(0) \exp(-p \zeta_T \tau)$, it proves
also convenient to consider the actual velocity field 
$\delta \mathbf{u}(\tau)= v_H(\tau) \mathbf{w}(\tau)$
instead of its dimensionless counterpart $\mathbf{w}$, since the
prediction (\ref{solucion_wp}) then takes the form :
\begin{equation}\label{solucion_up}
 \delta u_{\mathbf{k}\perp}(\tau)=
\exp\left\{-\frac{\tilde{\eta} \tilde{k}^2}{4p\zeta_n}
\left[\frac{n_H^2(0)}{n_H^2(\tau)}-1\right]
\right\}\delta u_{\mathbf{k}\perp}(0).
\end{equation} 
The plot of $\delta u_{\mathbf{k}\perp}(\tau)/\delta u_{\mathbf{k}\perp}(0)$
as a function of $(n_H(0)/n_H(\tau))^2$ allows then by simple
exponential fitting to extract  $\tilde{\eta}/\zeta_n$ (recall
that $\tilde{k}=l_H(0) k$ is known). 
Figure \ref{etas} compares the ratios $\tilde{\eta}/\zeta_n$
extracted from such fits for various values of $p$ with the
theoretical prediction. We have plotted the value of the 
viscosity normalized by its elastic value, $\tilde{\eta}_e$. For $p=0.1$ 
the agreement is quite
good but for $p=1$ we obtain discrepancies of the order of $15\%$.
Such deviations could be due to the limitation for high
dissipation of the first Sonine approximation that has been
used to compute the numerical values of the various quantities
involved in the description (in particular, the deviations of
the homogeneous decay state velocity distribution from its Gaussian form
might be relevant). Additionally, the
shear viscosity could suffer from finite size effects.
For a related discussion in the realm of granular gases, 
where both effects alluded to are at work, see \cite{msg07,gsm07,bgm08}. 
Finally, neglecting the $k^2$ contribution to  
$\delta \zeta_\mathbf{u}$ might not be innocuous.
From symmetry considerations, such a term must be of the form 
$k^2l_H(\tau)^2\mathbf{w}_\mathbf{k}$,
so that the equation (\ref{ec_l_u1t}) for the transversal velocity 
has the same form as the one we considered, but with a
``shifted'' shear viscosity. It is worth pointing out here that 
in the corresponding equation (\ref{solucion_wp}), putative order
$k^2$ corrections to $\delta \zeta_n$ and $\delta\zeta_T$ play no role
(see Eq. (\ref{ec_l_u1t})) : the decay rates $\zeta_n$ 
and $\zeta_T$ appearing in (\ref{solucion_wp}) are fingerprints
of the $\tau$ dependence of $l_H$ and of the rescaling procedure leading to 
$\mathbf{w}_\mathbf{k}$ from the actual velocity flow. Those two decay rates
are therefore properties of the homogeneous solution and do not 
suffer any  finite $k$ correction.

\subsection{Perturbation of the longitudinal velocity}

 \begin{figure}
\includegraphics[angle=0,width=0.46\textwidth]{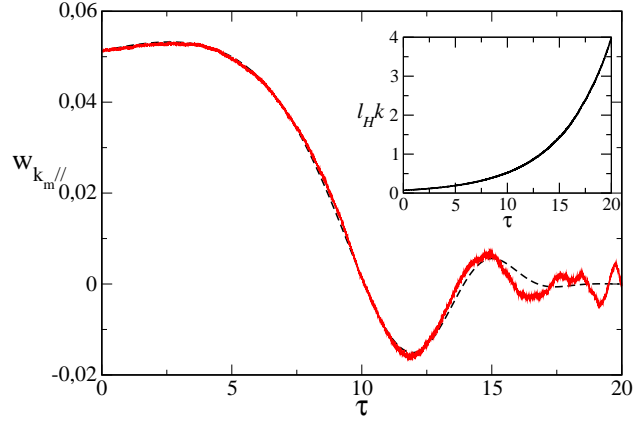}
\caption{Time evolution of the longitudinal velocity for a system with
  $p=0.1$ as a function of $\tau$. The solid line shows the simulation
  results and the dashed line is the numerical solution of
  (\ref{sistema}). For $\tau=15$ the rescaled density is 
  $n\sigma^2\simeq 2.5.10^{-3}$. The inset shows the increase of mean-free 
  path with time and that for $\tau >10$, $l_H k$ is no longer a small
  quantity.}\label{vparalelap0.1}
\end{figure}

\begin{figure}
\includegraphics[angle=0,width=0.46\textwidth]{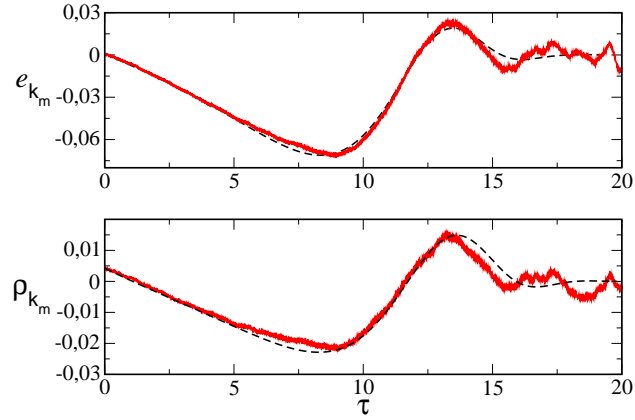}
\caption{Time evolution of $e_{\mathbf{k_m}}$ and
$\rho_{\mathbf{k_m}}$ as a function of $\tau$ for a system with
$p=0.1$. The solid lines show Molecular Dynamics results while the
dashed lines are for the numerical solution of (\ref{sistema}). For $\tau=15$ 
the rescaled density is $n\sigma^2\simeq 2.5.10^{-3}$.}
\label{enerdensip0.1}
\end{figure}

In order to investigate further the validity of the
hydrodynamic equations, we consider
a perturbation of the longitudinal velocity
\begin{equation}
u_x(\mathbf{r},0)=A\sin(k_{m}x),
\label{eq:pertlong}
\end{equation}
where $A=10^{-1}v_H(0)$ and $k_m=2\pi/L$. Since the hydrodynamic matrix
$M$ depends on time and $\frac{dM}{dt}$ does not commute with $M$, we
could not solve analytically the set of equations (\ref{sistema}), and
we turned to a numerical integration, using the transport coefficients
computed in Appendix \ref{appendixA}.

 \begin{figure}
\includegraphics[angle=0,width=0.46\textwidth]{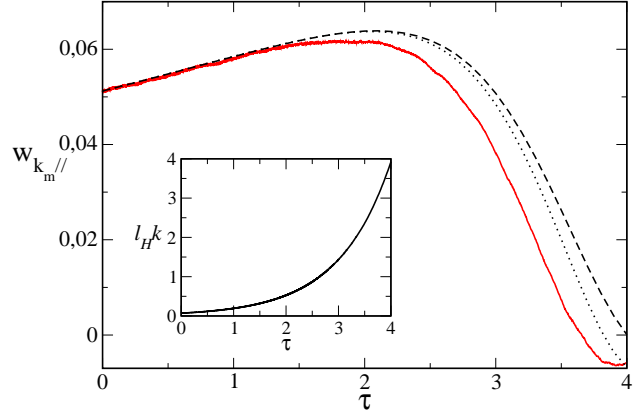}
\caption{Time evolution of the longitudinal velocity for a system with
  $p=0.5$ as a function of $\tau$. The solid line shows the simulation results
and the dashed line is for the numerical solution. We have also plotted 
with a dotted line the
numerical solution considering the elastic values of $\tilde{\kappa}$ and 
$\tilde{\mu}$ (i.e. their limit when $p\to 0^+$).
The inset shows the increase of mean-free path with time.
}\label{vparalelap0.5}
\end{figure}

\begin{figure}
\includegraphics[angle=0,width=0.46\textwidth]{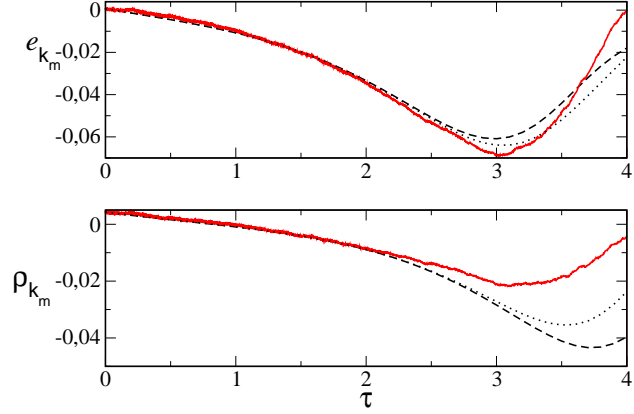}
\caption{Time evolution of $e_{\mathbf{k_m}}$ and $\rho_{\mathbf{k_m}}$ as a
function of $\tau$ for a system with $p=0.5$. The different lines
have the same meaning as in Fig. \ref{vparalelap0.5}.}\label{enerdensip0.5}
\end{figure}

In Fig. \ref{vparalelap0.1}, we have plotted the time evolution of the
rescaled longitudinal velocity field, $w_{k_{m}\parallel}$, as a
function of the internal time clock $\tau$ (number of collisions
per particle) for a system with
$p=0.1$. The results have been averaged over $16$ trajectories. It can
be seen that the theoretical framework is able to account for the
non trivial time dependence of the perturbation dynamics.
Moreover, as the equations for $\theta_{k_{m}}$ and $\rho_{k_{m}}$ are
coupled with the equation for $w_{k_{m}\parallel}$, a perturbation
such as (\ref{eq:pertlong}) induces a response of the above two other
fields (at variance with the transversal velocity, whose dynamics is
decoupled from the other three modes, at least at the linear level of
description adopted here). In Fig. \ref{enerdensip0.1}, we have
plotted the energy density
$e_{k_{m}}=\theta_{k_m}+\rho_{k_m}$ and $\rho_{k_{m}}$ as a
function of $\tau$. The agreement with theory is
very good both qualitatively and quantitatively, 
with an evolution that is well predicted till $\tau\simeq 15$ ;
for long times the simulation data become somewhat noisy (less
statistics can be achieved due to the smaller number of particles left
in the system. The number of particles at $\tau=15$ is $N\simeq 5000$).

In Fig.s \ref{vparalelap0.5} and \ref{enerdensip0.5}, results for a system
with $p=0.5$ are shown. Data are averaged over $64$ runs. 
In this case, the agreement is still good qualitatively, with
similar shapes for the theoretical and numerical curves,
but some discrepancies are observed.
The most significant deviation from the theoretical
prediction occurs
for the density, $\rho_{k_{m}}$, where the value of the minimum and
also its position are not predicted accurately. However, the
hydrodynamic framework still captures correctly the trends of the complex
dynamics of the perturbations. The above discrepancies could be
ascribable to the failure of the first Sonine approximation for high
dissipation (i.e. ``high'' $p$) or to finite size effects. We
mention here that within the first Sonine approximation, the
dimensionless coefficients $\tilde{\mu}$ and $\tilde{\kappa}$ exhibit
a divergent behavior in the vicinity of $p=0.8$ \cite{cdt04} which is
presumably unphysical, and is an indication of the limitation of the
method. For completeness and to assess the robustness of the
predictions with respect to a modification of the numerical values of
the key parameters, we have also reported in Fig.s
\ref{vparalelap0.5} and \ref{enerdensip0.5} the predictions obtained
when the transport coefficients take their elastic hard disc value
(e.g. $\tilde{\mu}(p=0.5)\simeq 0.505$ while $\tilde{\mu}(p=0)=0$, as
required by Fourier's law ). The important point is that the time
evolutions are not significantly affected, and that the main features
remain the same. Additionally, as mentioned after
Eq. (\ref{eq:resuapp}) and (\ref{solucion_up}), a possible source of
inaccuracy lies in the truncation of decay rates to their first order
($k^1$) in the gradients.  While the corresponding terms have been
shown to be small for inelastic hard spheres \cite{bdks98} (with the
notable simplification there that the velocity decay rate vanishes
identically due to momentum conservation), their relevance in the
present case has not been assessed, apart indirectly for the velocity
decay rate $\delta\boldsymbol{\zeta}_u$, by noting that second order
corrections do not spoil the accuracy of the prediction
(\ref{solucion_wp}), see Figures \ref{vtransversalp0.1} and
\ref{vtransversalp0.5}.

\section{Conclusions}

The objective here has been to explore the validity of a hydrodynamic
description based on density, momentum, and kinetic temperature fields
for a gas composed of particles which annihilate with probability $p$
or scatter elastically otherwise, by a direct analysis of the spectrum
of the linear Boltzmann equation.  The motivation mainly was to study
the applicability of hydrodynamics to systems which {\it a priori}
lack scale separation and in which there are no collisional
invariants.  The analysis performed here has been shown to lead to
results equivalent to those obtained previously from the more formal
Chapman-Enskog expansion \cite{cdt04}, with the difference that
the present approach considers linear excitations only.  However, the
current spectral method is arguably more straightforward and
explicitly shows that the hydrodynamic description arises in the
appropriate time scale, when the ``kinetic modes'' can be considered
as negligible against the hydrodynamic ones.

The eigenvalue problem of the Boltzmann operator linearized around the
homogeneous decay state has been addressed and we have identified the
hydrodynamic eigenfunctions. These eigenfunctions are not simply
linear combinations of $1$, $\mathbf{v}$ and $v^2$ as it happens in
the elastic case, but they are replaced by derivatives of the
homogeneous decay state velocity distribution function $\chi_H$, that
is not known analytically.  As a consequence of the non-hermitian
character of the linearized Boltzmann operator, the eigenfunctions are
not orthogonal. It is nevertheless possible to construct a set of
biorthonormal functions, $\{\bar{\xi}_\beta\}_{\beta=1,\ldots 3}$,
which are linear combinations of $1$, $\mathbf{v}$ and $v^2$, a
crucial point in order to obtain the hydrodynamic equations. The
analysis is complicated by the fact that none of the
$\{\bar{\xi}_\beta\}$ functions are left eigenfunctions of the
linearized Boltzmann operator, since no quantity is conserved during
binary encounters.  We have used these hydrodynamic eigenfunctions to
derive to Navier-Stokes order the heat and momentum fluxes, together
with the various decay rates. To this end, we have decomposed the
distribution function, $\delta\chi_\mathbf{k}$, into its hydrodynamic
and non-hydrodynamic parts. This decomposition enables us to close the
hydrodynamic equations in the long time limit and to order $k^2$, and
provides Green-Kubo formulas for the transport coefficients. We then
arrived at the linearized equations (around the homogeneous decay
state) for the hydrodynamic fields, in the usual form of partial
differential equations with coefficients that are independent of the
space variable but depend on time, since the reference state
considered is itself time dependent. If we analyze the stability of
these equations, we may conclude that small perturbations should decay
in the long time limit. Nevertheless, it must be stressed that the
perturbation may increase at short times, thereby possibly leaving
the linear domain where our analysis holds. The long time dynamics in
such a case remains an open question.

In section \ref{simulaciones}, we have reported Molecular Dynamics
simulations for the evolution of a perturbation of the transversal and
longitudinal velocity fields, which show a rich dynamics.  The
agreement between theory and simulations is very good for moderate values
of the annihilation probability $p$ (say $p<0.5$), which gives strong
support to the theoretical analysis developed here. The theoretical
curves still agree qualitatively at larger $p$, with however some
quantitative discrepancies which might be a
manifestation of the approximations underlying the computation of the
transport coefficients (namely $\kappa$ and $\mu$, evaluated to first
order in a Sonine expansion \cite{cdt04}). Indeed, those coefficients
are predicted to exhibit a divergent behavior for $p\sim 0.8$, but
the simulations we have performed do not show a qualitatively
different behavior for these values of $p$.  A further complication
--that is also an {\it a priori} limitation for the efficiency of a
hydrodynamic approach-- is that for values of $p$ close to unity, the
separation of time scales between the kinetic and hydrodynamic modes
is not clear cut (the density decay rate is on the order of the
collision frequency, comparable to the inverse typical time of the
kinetic modes).  To address this concern, one should study in detail
the spectrum of non hydrodynamic modes to find the slowest, and
compare it to the fastest decay rate in our problem (i.e. $\zeta_n$).
Such a program, left for future work, has been achieved in Appendix 
\ref{appendixC} within the Maxwell model framework, with the
conclusion that scale separation does not hold for $p>p^*=1/4$. 
While the threshold $p^*$  is {\it a priori} model dependent,
a similar phenomenon is to be expected in the original ``hard-sphere''
dynamics considered in this paper. The coarse-grained description 
for $p>p^*$ is an open question.

\begin{acknowledgments}
We would like to thank the Agence Nationale de la Recherche
for financial support. M.~I.~G.~S.~acknowledges financial support from Becas de
la 
Fundaci\'on La Caixa y el Gobierno Franc\'es and from the HPC-EUROPA project (RII3-CT-2003-506079),
with the support of the European Community Research Infrastructure Action.

\end{acknowledgments}

\begin{appendix}

\section{Approximate expression for the transport coefficients and decay rates}
\label{appendixA}
For completeness, we recall here the explicit expressions used for the transport coefficients and
decay rates, as obtained within the first order Sonine scheme in 
Ref. \cite{cdt04}. 
The distribution function in this approximation reads
\begin{equation}
\chi_H(c)=\frac{1}{\pi^{d/2}}e^{-c^2}\left\{1
+a_2\left[\frac{d(d+2)}{8}-\frac{d+2}{2}c^2+\frac{c^4}{2}\right]\right\},
\end{equation}
with the kurtosis
\begin{equation}
a_2=\frac{8(3-2\sqrt{2})p}{(4d+6-\sqrt{2})p+8\sqrt{2}(d-1)(1-p)}.
\end{equation}
This expression  allows us to calculate the
decay rates
\begin{eqnarray}
\zeta_n&=&\frac{d+2}{4}\left(1-a_2\frac{1}{16}\right),\\
\zeta_T&=&\frac{d+2}{8d}\left(1+a_2\frac{8d+11}{16}\right),
\end{eqnarray}
and also the transport coefficients
\begin{equation}
\tilde{\eta}=\frac{1}{4\nu_\eta-2p\zeta_T},
\end{equation}
\begin{equation}
\tilde{\kappa}=\frac{1}{\nu_\kappa-2p\zeta_T}
\left[\frac{1}{2}p\zeta_n\tilde{\mu}+\frac{d+2}{8}(2a_2+1)\right],
\end{equation}
\begin{equation}
\tilde{\mu}=\frac{1}{2\nu_\kappa-3p\zeta_T-2p\zeta_n}
\left[p\zeta_T\tilde{\kappa}+\frac{d+2}{8}(2a_2+1)\right],
\end{equation}
where the values of the coefficients $\nu_\eta$ and $\nu_\kappa$ are
\begin{equation}
\nu_\eta=
\frac{p}{8d}\left[3+6d+2d^2-a_2\frac{278+375d+96d^2+2d^3}{32(d+2)}\right]
+(1-p)\left(1-a_2\frac{1}{32}\right),
\end{equation}
\begin{equation}
\nu_\kappa=\frac{p}{32d}
\left[16+27d+8d^2+a_2\frac{2880+1544d-2658d^2-1539d^3-200d^4}{32d(d+2)}\right]
+(1-p)\frac{d-1}{d}\left(1+a_2\frac{1}{32}\right).
\end{equation}
Finally, the expressions for $\zeta_{u,\rho}$ and 
$\zeta_{u,\theta}$ are
\begin{eqnarray}
\zeta_{u,\rho}&=&\frac{8(d-1)}{d(d+2)}\tilde{\mu}\zeta_u,\\
\zeta_{u,\theta}&=&\frac{8(d-1)}{d(d+2)}\tilde{\kappa}\zeta_u,
\end{eqnarray}
with
\begin{equation}
\zeta_u=\frac{(d+2)^2}{32(d-1)}
\left[1+a_2\frac{-86-101d+32d^2+88d^3+28d^4}{32(d+2)}\right].
\end{equation}

\section{Eigenfunctions of $\Lambda$}\label{appendixB}
In this Appendix some of the details leading to the solution of the
eigenvalue problem (\ref{problema_autov}) are given.
Consider first the function
\begin{equation}
\Psi_1(\mathbf{c})=\chi_H(\mathbf{c}).
\end{equation}
and let the linearized operator $\Lambda$ act onto $\chi_H$
\begin{eqnarray}\label{ap:c2}
\Lambda(\mathbf{c}_1)\chi_H(\mathbf{c}_1)&=&\gamma\int\!\!d\mathbf{c}_2
T(\mathbf{c}_1,\mathbf{c}_2)(1+{\cal P}_{12})
\chi_H(\mathbf{c}_1)\chi_H(\mathbf{c}_2)
\nonumber\\
&+&p(2\zeta_n-d\zeta_T)\chi_H(\mathbf{c}_1)-p\zeta_T\mathbf{c}_1\cdot
\frac{\partial}{\partial\mathbf{c}_1}\chi_H(\mathbf{c}_1).
\end{eqnarray}
If we take into account the equation for $\chi_H(\mathbf{c}_1)$
\begin{equation}\label{ap:c3}
(d\zeta_T-2\zeta_n)p\chi_H(\mathbf{c}_1)+p\zeta_T\mathbf{c}_1\cdot
\frac{\partial}{\partial\mathbf{c}_1}\chi_H(\mathbf{c}_1)=\gamma\int\!\!
d\mathbf{c}_2 T(\mathbf{c}_1,\mathbf{c}_2)
\chi_H(\mathbf{c}_1)\chi_H(\mathbf{c}_2),
\end{equation}
we can rewrite equation (\ref{ap:c2}) as
\begin{equation}\label{ap:c4}
\Lambda(\mathbf{c}_1)\chi_H(\mathbf{c}_1)
=(d\zeta_T-2\zeta_n)p\chi_H(\mathbf{c}_1)
+p\zeta_T\mathbf{c}_1\cdot\frac{\partial}{\partial\mathbf{c}_1}
\chi_H(\mathbf{c}_1).
\end{equation}

Consider now the function 
\begin{equation}\label{ap:c5}
\Psi_2(\mathbf{c})=\mathbf{c}\cdot
\frac{\partial}{\partial\mathbf{c}}\chi_H(\mathbf{c}).
\end{equation}
In order to proceed, we perform the change of variables
$\mathbf{c}_1=\lambda\mathbf{c}_1^\prime$ in equation (\ref{ap:c3})
\begin{eqnarray}
(d\zeta_T&-&2\zeta_n)p\chi_H(\lambda\mathbf{c}_1^\prime)
+p\zeta_T\mathbf{c}_1^\prime\cdot
\frac{\partial}{\partial\mathbf{c}_1^\prime}
\chi_H(\lambda\mathbf{c}_1^\prime)\nonumber\\
&=&\gamma
\lambda^{d+1}\int\!\!d\mathbf{c}_2^\prime\!\!\int\!\!
d\hat{\boldsymbol{\sigma}}\theta(\mathbf{c}_{12}^\prime\cdot
\hat{\boldsymbol{\sigma}})\mathbf{c}_{12}^\prime\cdot
\hat{\boldsymbol{\sigma}}[(1-p)b_{\boldsymbol{\sigma}}^{-1}-1]
\chi_H(\lambda\mathbf{c}_1^\prime)\chi_H(\lambda\mathbf{c}_2^\prime).
\nonumber\\
\end{eqnarray}
Deriving with respect to $\lambda$ we obtain
\begin{eqnarray}
(d\zeta_T-2\zeta_n)p\frac{\partial}{\partial\lambda}
\chi_H(\lambda\mathbf{c}_1)
&+&p\zeta_T\mathbf{c}_1\cdot\frac{\partial}{\partial\mathbf{c}_1}
\left(\frac{\partial}{\partial\lambda}
\chi_H(\lambda\mathbf{c}_1)\right)\nonumber\\
&=&(d+1)\lambda^d\gamma\int\!\!d\mathbf{c}_2
T(\mathbf{c}_1,\mathbf{c}_2)\chi_H(\lambda\mathbf{c}_1)
\chi_H(\lambda\mathbf{c}_2)\nonumber\\
&+&\gamma\lambda^{d+1}\int\!\!d\mathbf{c}_2
T(\mathbf{c}_1,\mathbf{c}_2)\frac{\partial\chi_H(\lambda\mathbf{c}_1)}
{\partial\lambda}\chi_H(\lambda\mathbf{c}_2)\nonumber\\
&+&\gamma\lambda^{d+1}\int\!\!d\mathbf{c}_2T(\mathbf{c}_1,\mathbf{c}_2)
\chi_H(\lambda\mathbf{c}_1)
\frac{\partial\chi_H(\lambda\mathbf{c}_2)}{\partial\lambda}.
\end{eqnarray}
and taking $\lambda=1$ we arrive at the equation for $\Psi_2(\mathbf{c})$
\begin{eqnarray}\label{ap:c7}
(d\zeta_T-2\zeta_n)p\Psi_2(\mathbf{c}_1)&+&p\zeta_T\mathbf{c}_1\cdot
\frac{\partial}{\partial\mathbf{c}_1}\Psi_2(\mathbf{c}_1)\nonumber\\
&=&(d+1)\gamma\int\!\!d\mathbf{c}_2
T(\mathbf{c}_1,\mathbf{c}_2)\chi_H(\lambda\mathbf{c}_1)
\chi_H(\lambda\mathbf{c}_2)\nonumber\\
&+&\gamma\int\!\!d\mathbf{c}_2
T(\mathbf{c}_1,\mathbf{c}_2)\Psi_2(\mathbf{c}_1)
\chi_H(\lambda\mathbf{c}_2)\nonumber\\
&+&\gamma\int\!\!d\mathbf{c}_2T(\mathbf{c}_1,\mathbf{c}_2)
\chi_H(\lambda\mathbf{c}_1)\Psi_2(\mathbf{c}_2),
\end{eqnarray}
or equivalently
\begin{equation}\label{ap:c8}
\Lambda(\mathbf{c}_1)\Psi_2(\mathbf{c}_1)=-(d+1)\left[(d\zeta_T-2\zeta_n)p
\chi_H(\mathbf{c}_1)+p\zeta_T\mathbf{c}_1\cdot
\frac{\partial}{\partial\mathbf{c}_1}\chi_H(\mathbf{c}_1)\right].
\end{equation}
Consequently, equations (\ref{ap:c4}) and (\ref{ap:c8}) can be written as
\begin{eqnarray}
\Lambda(\mathbf{c}_1)\Psi_1(\mathbf{c}_1)
&=&(d\zeta_T-2\zeta_n)p\Psi_1(\mathbf{c}_1)+p\zeta_T\Psi_2(\mathbf{c}_1),
\label{ap:c9}\\
\Lambda(\mathbf{c}_1)\Psi_2(\mathbf{c}_2)
&=&-(d+1)[(d\zeta_T-2\zeta_n)p\Psi_1(\mathbf{c}_1)+p\zeta_T
\Psi_2(\mathbf{c}_1)].\label{ap:c10}
\end{eqnarray}
With equations (\ref{ap:c9}) and (\ref{ap:c10}) we can easily see that
\begin{equation}
\Lambda(\mathbf{c}_1)[(d+1)\Psi_1(\mathbf{c}_1)+\Psi_2(\mathbf{c}_1)]=0,
\end{equation}
so  that
\begin{equation}
\xi_1(\mathbf{c}_1)\equiv(d+1)\Psi_1(\mathbf{c}_1)+\Psi_2(\mathbf{c}_1),
\end{equation}
is an eigenfunction of the operator $\Lambda(\mathbf{c}_1)$ with eigenvalue
$\lambda_1=0$. It is also straightforward to see that
\begin{equation}
\xi_2(\mathbf{c}_1)=(d\zeta_T-2\zeta_n)p\Psi_1(\mathbf{c}_1)
+p\zeta_T\Psi_2(\mathbf{c}_1),
\end{equation}
is an eigenfunction of $\Lambda$. We obtain that
\begin{eqnarray}
\Lambda(\mathbf{c}_1)\xi_2(\mathbf{c}_1)
&=&(d\zeta_T-2\zeta_n)p\Lambda(\mathbf{c}_1)\Psi_1(\mathbf{c}_1)
+p\zeta_T\Lambda(\mathbf{c}_1)\Psi_2(\mathbf{c}_1)\nonumber\\
&=&-(\zeta_T+2\zeta_n)p\xi_2(\mathbf{c}_1),
\end{eqnarray} 
where we have used equations (\ref{ap:c9}) and (\ref{ap:c10}). Therefore, we have 
that $\xi_2(\mathbf{c}_1)$ is an eigenfunction of $\Lambda(\mathbf{c}_1)$ with 
eigenvalue $\lambda_2=-(\zeta_T+2\zeta_n)p$.

Finally, let us consider the last function
\begin{equation}
\boldsymbol{\Psi}_3(\mathbf{c})
=-\frac{\partial}{\partial\mathbf{c}}\chi_H(\mathbf{c}).
\end{equation}
Deriving the equation obeyed by 
$\chi_H(\mathbf{c}-\mathbf{w})$, with respect to 
$\mathbf{w}$ and subsequently evaluating 
the result for $\mathbf{w}=\mathbf{0}$, we obtain


\begin{eqnarray}
(d\zeta_T-2\zeta_n)p\boldsymbol{\Psi}_3(\mathbf{c}_1)
&+&p\zeta_T\boldsymbol{\Psi}_3(\mathbf{c}_1)+p\zeta_T\mathbf{c}_1\cdot
\frac{\partial}{\partial\mathbf{c}_1}\boldsymbol{\Psi}_3(\mathbf{c}_1)
\nonumber\\
&=&\gamma\int\!\!d\mathbf{c}_2 T(\mathbf{c}_1,\mathbf{c}_2)
(1+{\cal P}_{12})\boldsymbol{\Psi}_3(\mathbf{c}_1)\chi_H(\mathbf{c}_2),
\end{eqnarray}
or equivalently
\begin{equation}
\Lambda(\mathbf{c}_1)\boldsymbol{\Psi}_3(\mathbf{c}_1)
=p\zeta_T\boldsymbol{\Psi}_3(\mathbf{c}_1).
\end{equation}
In other words, 
$\boldsymbol{\xi}_3(\mathbf{c}_1)\equiv\boldsymbol{\Psi}_3(\mathbf{c}_1)$ is
an eigenfunction of $\Lambda(\mathbf{c}_1)$ with eigenvalue 
$\lambda_3=p\zeta_T$.

\section{Linearized Boltzmann operator for Maxwell 
Molecules}\label{appendixC}
The objective in this Appendix is to study the spectrum of the linearized 
Boltzmann operator for Maxwell
molecules with annihilation. It will be shown that for 
$0\leq p\leq p^*$, where $p^*$ depends on the specific Maxwell model under
consideration, the norm of the hydrodynamic eigenvalues are smaller than the 
rest of the spectrum.

The main characteristic of Maxwell models is that the differential cross
section multiplied by the relative velocity is independent of the relative 
velocity. We are going to assume that
it is also independent of the angle between the two colliding particles. Then,
the Boltzmann equation for a system of Maxwell molecules which annihilate in a
collision with probability $p$ and collide elastically otherwise (with
probability $1-p$) reads
\begin{eqnarray}
\left(\frac{\partial}{\partial t}
+\mathbf{v}_1\cdot\nabla\right)f(\mathbf{r},\mathbf{v}_1,t)&=&-p\beta\Omega
\int\!\!d\mathbf{v}_2f(\mathbf{r},\mathbf{v}_1,t)f(\mathbf{r},\mathbf{v}_2,t)
\nonumber\\
&+&(1-p)\beta\int\!\!d\mathbf{v}_2\!\!\int\!\!d\hat{\boldsymbol{\sigma}}
[b_{\sigma}^{-1}-1]f(\mathbf{r},\mathbf{v}_1,t)f(\mathbf{r},\mathbf{v}_2,t),
\nonumber\\  
\end{eqnarray}
where $\beta$ is a constant representing the microscopic scattering collision
frequency, $\Omega=2\pi^{d/2}/\Gamma(d/2)$ is
the $d$-dimensional solid angle and the operator $b_{\sigma}^{-1}$ is defined
in the main text, Eq. (\ref{op_bmenos1}). 

If we consider the homogeneous case, it is straightforward to see that the
temperature is a constant in time, $T_H(t)=T_H(0)$, and that the density
evolves as
\begin{equation}\label{densityMax}
n_H(t)=\frac{n_H(0)}{1+p\zeta_nt}, 
\end{equation}
with $\zeta_n=\beta\Omega n_H(0)$. Moreover,
it was shown in \cite{commentsb85} that there is an exact mapping between the
homogeneous equation for Maxwell molecules with annihilation (arbitrary $p$) 
and the usual elastic  Maxwell molecules ($p=0$). If we introduce the time
scale 
\begin{equation}\label{escalasMax}
s(t)=\int_0^tdt'\frac{n_H(t')}{n_H(0)}, 
\end{equation}
it can be seen that the distribution function for arbitrary $p$ is
\begin{equation}
f(\mathbf{v},t)=\frac{n_H(t)}{n_H(0)}f^E\left[\mathbf{v},(1-p)s(t)\right],
\end{equation}
where $n_H(t)$ is given by formula (\ref{densityMax}), $s(t)$ by 
(\ref{escalasMax}), and the function $f^E$ is the
distribution function for elastic Maxwell molecules. Note that this relation
is also valid for $p=1$ where $f^E$ is frozen in the initial condition.

In the elastic case, every homogeneous distribution 
tends to relax to a Maxwellian after a transient time. Then, the same is going
to happen for arbitrary $p<1$ because of the mapping. In this
sense, we can consider that the state analogous to the homogeneous decay state introduced in the 
main text for hard particles and that will constitute the appropriate
reference  will be characterized by the distribution function
\begin{equation}\label{fM}
f_H(\mathbf{v},t)=\frac{n_H(t)}{v_H^d}\chi_M(v/v_H), 
\end{equation}
where $v_H=\left(\frac{2T_H}{m}\right)^{1/2}$ and
$\chi_M(v)=1/\pi^{d/2}e^{-v^2}$ is the Maxwellian distribution. Note that in 
the homogeneous decay state, both the density and temperature decay, whether in the present
case only the density decays. Then, as $v_H$ plays no role, we will consider 
units with $v_H=1$ for simplicity. 

Let us study now the linear response to an inhomogeneous small perturbation
around the reference state as it was done in section \ref{sec:boltz}. 
If we introduce the scaled distribution function
\begin{equation}
\delta\chi(\mathbf{r},\mathbf{v},\tau)=\frac{n_H(0)}{n_H(t)}
[f(\mathbf{r},\mathbf{v},t)-f_H(\mathbf{v},t)],
\end{equation}
the equation for $\delta\chi$ in the $s$ scale defined in (\ref{escalasMax}) is
\begin{equation}\label{lbem}
\left[\frac{\partial}{\partial s}
+h_H(s)\mathbf{v}_1\cdot\nabla\right]
\delta\chi(\mathbf{r},\mathbf{v},s)=\Lambda(\mathbf{v}_1)
\delta\chi(\mathbf{r},\mathbf{v},s),
\end{equation}
where we have introduced the function $h_H(s)=n_H(0)/n_H(s)$, and the 
linearized Boltzmann operator
\begin{eqnarray}
\Lambda(\mathbf{v}_1)g(\mathbf{v}_1)&=&(1-p)\frac{\zeta_n}{\Omega}
\int\!\!d\mathbf{v}_2\!\!\int\!\!d\hat{\boldsymbol{\sigma}}
[b_{\sigma}^{-1}-1](1+{\cal P}_{12})\chi_M(v_1)g(\mathbf{v}_2)\nonumber\\
&-&p\zeta_n\chi_M(v_1)
\int\!\!d\mathbf{v}_2g(\mathbf{v}_2).
\end{eqnarray} 
Let us stress that, although there is an exact mapping for the full
non-linear homogeneous equation between $p=0$ and arbitrary $p$, 
no such mapping exists for the linear inhomogeneous Boltzmann equation, 
Eq. (\ref{lbem}). Then, as in the main text, the possibility of an
hydrodynamic description depends on the properties of the linearized Boltzmann
operator. Here we will see that it is possible to calculate all the
eigenfunctions and eigenvalues of this operator and that there is a region of
the parameter $p$ in which we have an appropriate scale separation.

Let us write the linearized Boltzmann operator as
\begin{equation}\label{desocom_lambda}
\Lambda(\mathbf{v}_1)g(\mathbf{v}_1)=(1-p)\Lambda^E(\mathbf{v}_1)
g(\mathbf{v}_1)-p\zeta_n\chi_M(v_1)\int\!\!d\mathbf{v}_2g(\mathbf{v}_2),
\end{equation}
where we have introduced the linearized Boltzmann operator for elastic Maxwell 
particles
\begin{equation}
\Lambda^E(\mathbf{v}_1)=\frac{\zeta_n}{\Omega}\int\!\!d\mathbf{v}_2
\!\!\int\!\!d\hat{\boldsymbol{\sigma}}[b_{\sigma}^{-1}-1]
(1+{\cal P}_{12})\chi_M(v_1)g(\mathbf{v}_2),
\end{equation}
whose spectral properties are well known \cite{resibois,mclennan}. In
particular, for $d=3$ its eigenfunctions are
\begin{equation}
\phi_{rlm}(\mathbf{v})=A_{rl}\chi_M(v)S^r_{l+1/2}(v^2)v^l
Y_{lm}(\theta,\varphi),\qquad r=0,1,\dots
\end{equation}
where $Y_{lm}(\theta,\varphi)$ are the spherical harmonics, functions of the
polar angles $(\theta,\varphi)$ of $\mathbf{v}$ with respect to an arbitrary
direction,  $S^r_{l+1/2}(v^2)$ are the Sonine polynomials which satisfy
\begin{equation}\label{sonine_properties}
\int_0^{\infty}dxe^{-x}S^n_{q}(x)S^{n'}_{q}(x)
=\frac{\Gamma(n+q+1)}{n!}\delta_{nn'},
\end{equation}
and $A_{rl}$ are some constants that are introduced in order to normalize the 
eigenfunctions and that play no role in the following analysis. The eigenvalues
of $\Lambda^E$, $\lambda^E_{rl}$, are also known. It can be seen that
$\lambda^E_{00}$, $\lambda^E_{10}$, and $\lambda^E_{01}$ (which is 3 times degenerate) vanish, corresponding 
to the five hydrodynamic eigenvalues and 
that the slowest kinetic mode corresponds to an eigenvalue $\lambda^E_k=-\zeta_n/3$.

\begin{figure}
\includegraphics[angle=0,width=0.46\textwidth]{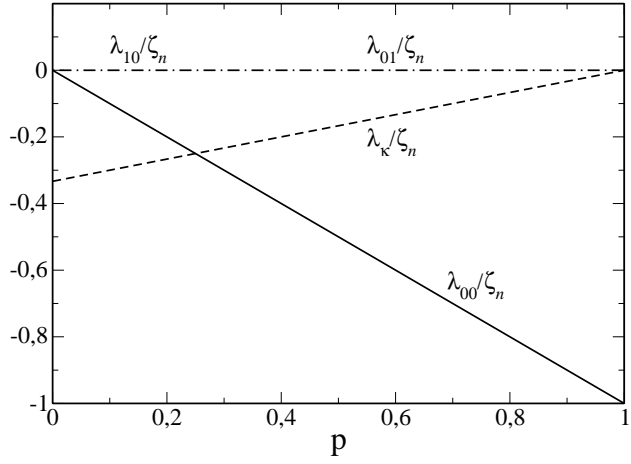}
\caption{Hydrodynamic eigenvalues, $\lambda_{00}$, $\lambda_{10}$, and 
$\lambda_{01}$, and the slowest kinetic eigenvalue, $\lambda_k$, as a function 
of the dissipation parameter $p$. The eigenvalues are normalized by
$\zeta_n=\beta \Omega n_H(0)$.}\label{autovalores}
\end{figure}

The important point here is that the functions $\phi_{rlm}$ are also
eigenfunctions of the second operator in (\ref{desocom_lambda}). Taking into
account the orthogonality properties of the spherical harmonics and of the
Sonine polynomials, Eq. (\ref{sonine_properties}), one has 
\begin{equation}\label{autovalores2}
\int\!\!d\mathbf{v}\phi_{rlm}(\mathbf{v})
=\delta_{r0}\delta_{l0}.
\end{equation}
Then, as $\phi_{000}(\mathbf{v})=\chi_M(v)$, the functions 
$\phi_{rlm}(\mathbf{v})$ are in fact the eigenfunctions
of the total linearized Boltzmann operator for arbitrary $p$. With the aid of
(\ref{autovalores2}) is straightforward to see that the eigenvalues are
\begin{equation}
\lambda_{rl}=(1-p)\lambda^E_{rl}-p\zeta_n\delta_{r0}\delta_{l0}.
\end{equation}
In Fig. \ref{autovalores} we plot the hydrodynamic eigenvalues as well as the 
slowest kinetic
eigenvalue, $\lambda_k$, as a function of the dissipation parameter $p$. 
It can be seen that for
$0\leq p<1/4$, there is scale separation in the sense that 
the three modes (density, linear momentum and kinetic energy) retained in the
coarse-grained description decay slower that any of the other ``kinetic'' modes.
On the other hand, for $1/4<p\leq 1$, the largest kinetic
eigenvalue is slower than $\lambda_{00}$. We therefore conclude here
that a conservative requirement for the validity of our approach in the case
of Maxwell molecules would be $p<1/4$.

\section{Evaluation of $\delta\zeta_n$,
$\delta\boldsymbol{\zeta}_u$ and
$\delta\zeta_T$}
\label{appendixD}
In this Appendix we calculate the contribution of the
hydrodynamic part of $\delta\chi_\mathbf{k}$ to the functionals 
$\delta\zeta_n$,
$\delta\boldsymbol{\zeta}_u$ and
$\delta\zeta_T$ defined in (\ref{def_zeta_n})-(\ref{def_zeta_T}). To this end,
we write explicitly $P\delta\chi_\mathbf{k}$ as
\begin{eqnarray}
P\delta\chi_\mathbf{k}(\mathbf{c}_1)&=&
\rho_\mathbf{k}\xi_1(\mathbf{c}_1)+\left(\frac{1}{2}\theta_\mathbf{k}
+\rho_\mathbf{k}\right)\xi_2(\mathbf{c}_1)
+\mathbf{w}_\mathbf{k}\cdot\boldsymbol{\xi}_3(\mathbf{c}_1)\nonumber\\
&=&\rho_\mathbf{k}\chi_H(\mathbf{c}_1)-\frac{1}{2}\theta_\mathbf{k}
\frac{\partial}{\partial\mathbf{c}_1}\cdot[\mathbf{c}_1\chi_H(\mathbf{c}_1)]
-\mathbf{w}_\mathbf{k}\cdot\frac{\partial}{\partial\mathbf{c}_1}
\chi_H(\mathbf{c}_1).
\end{eqnarray}

Let us first evaluate $\delta\zeta_n[P\delta\chi_{\mathbf{k}}]$. After some
algebra it can be seen that
\begin{eqnarray}
\delta\zeta_n[\chi_H(\mathbf{c}_1)]&=&-4\zeta_n,\label{ap:e2}\\
\delta\zeta_n\left[\frac{\partial}{\partial\mathbf{c}_1}
\cdot[\mathbf{c}_1\chi_H(\mathbf{c}_1)]\right]&=&2\zeta_n,\label{ap:e3}\\
\delta\zeta_n\left[\frac{\partial}{\partial{\mathbf{c}_1}}
\chi_H(\mathbf{c}_1)\right]&=&0,\label{ap:e4}
\end{eqnarray}
where we have used equations (\ref{ap:c5}), (\ref{ap:c7}), the definition of
the density decay rate, equation (\ref{zeta_n}), and symmetry 
considerations. Then, if we consider equations (\ref{ap:e2})-(\ref{ap:e4}) we 
finally obtain
\begin{equation}
\delta\zeta_n[P\delta\chi_{\mathbf{k}}]=-4\zeta_n\rho_\mathbf{k}
-\zeta_n\theta_\mathbf{k}.
\end{equation}

Now let us calculate
$\delta\boldsymbol{\zeta}_u[P\delta\chi_{\mathbf{k}}]$. Using the 
definition of the density decay rate, equation (\ref{zeta_n}), and symmetry 
considerations, it appears that
\begin{eqnarray}
\delta\boldsymbol{\zeta}_u[\chi_H(\mathbf{c}_1)]&=&0,\\
\delta\boldsymbol{\zeta}_u\left[\frac{\partial}{\partial\mathbf{c}_1}
\cdot(\mathbf{c}_1\chi_H(\mathbf{c}_1))\right]&=&0,\\
\delta\zeta_{u_i}\left[\frac{\partial}{\partial c_{1j}}\chi_H(\mathbf{c}_1)
\right]&=&\delta_{ij}\delta\zeta_{u_i}\left[\frac{\partial}{\partial c_{1i}}
\chi_H(\mathbf{c}_1)\right]=2\zeta_n.
\end{eqnarray}
Then, we have
\begin{equation}
\delta\boldsymbol{\zeta}_u[P\delta\chi_{\mathbf{k}}]
=-2\zeta_n\mathbf{w}_\mathbf{k}.
\end{equation}

Finally, we turn to
$\delta\zeta_T[P\delta\chi_{\mathbf{k}}]$. Using the definitions of
the decay rates, equations (\ref{zeta_n}) and (\ref{zeta_T}), we obtain
\begin{eqnarray}
\delta\zeta_T[\chi_H(\mathbf{c}_1)]&=&-4\zeta_T,\\
\delta\zeta_T\left[\frac{\partial}{\partial\mathbf{c}_1}
\cdot(\mathbf{c}_1\chi_H(\mathbf{c}_1))\right]&=&6\zeta_T+4\zeta_n,\\
\delta\zeta_T\left[\frac{\partial}{\partial{\mathbf{c}_1}}
\chi_H(\mathbf{c}_1)\right]&=&0
\end{eqnarray}
from which it follows that
\begin{equation}
\delta\zeta_T[P\delta\chi_{\mathbf{k}}]
=-4\zeta_T\rho_\mathbf{k}-(3\zeta_T+2\zeta_n)\theta_\mathbf{k}.
\end{equation}


\end{appendix}

\end{document}